\documentclass[twocolumn]{aastex63}
\usepackage{multirow}

\usepackage{placeins}
\FloatBarrier
\submitjournal{AJ}
\usepackage{float}
\usepackage[ansinew]{inputenc}
\usepackage[american]{babel}
\usepackage{etex}
\usepackage{stackengine}
\usepackage{csquotes}
\usepackage{ifthen}
\usepackage{url}
\usepackage{euscript}

\shorttitle{Host star metallicity of directly imaged wide-orbit planets: implications for planet formation}

\shortauthors{C. Swastik et al.}

\graphicspath{{./}{figures/}}
\usepackage{float}
\usepackage{natbib}
\begin{document}

\title{Host star metallicity of directly imaged wide-orbit planets: implications for planet formation}

\correspondingauthor{C. Swastik}
\email{swastik.chowbay@iiap.res.in}

\author[0000-0003-1371-8890]{C. Swastik}
\affiliation{Indian Institute of Astrophysics, Koramangala 2nd Block, Bangalore 560034, India}

\author[0000-0003-0799-969X]{Ravinder K. Banyal}
\affiliation{Indian Institute of Astrophysics, Koramangala 2nd Block, Bangalore 560034, India}

\author[0000-0002-0554-1151]{Mayank Narang}
\affiliation{Department of Astronomy and Astrophysics, Tata Institute of Fundamental Research
Homi Bhabha Road, Colaba, Mumbai 400005, India}

\author[0000-0002-3530-304X]{P. Manoj}
\affiliation{Department of Astronomy and Astrophysics, Tata Institute of Fundamental Research
Homi Bhabha Road, Colaba, Mumbai 400005, India}

\author{T. Sivarani}
\affiliation{Indian Institute of Astrophysics, Koramangala 2nd Block, Bangalore 560034, India}

\author{Bacham E. Reddy}
\affiliation{Indian Institute of Astrophysics, Koramangala 2nd Block, Bangalore 560034, India}

\author{S. P. Rajaguru}
\affiliation{Indian Institute of Astrophysics, Koramangala 2nd Block, Bangalore 560034, India}

\begin{abstract}
Directly imaged planets are self-luminous companions of pre-main sequence and young main sequence stars. They reside in wider orbits ($\sim10\mathrm{s}-1000\mathrm{s}$~AU) and generally are more massive compared to the close-in ($\lesssim 10$~AU) planets. Determining the host star properties of these outstretched planetary systems is important to understand and discern various planet formation and evolution scenarios. We present the stellar parameters and metallicity ([Fe/H]) for a subsample of 18 stars known to host planets discovered by the direct imaging technique.  We retrieved the high resolution spectra for these stars from public archives and used the synthetic spectral fitting technique and Bayesian analysis to determine the stellar properties in a uniform and consistent way. For eight sources, the metallicities are reported for the first time, while the results are consistent with the previous estimates for the other sources. Our analysis shows that metallicities of stars hosting directly imaged planets are close to solar with a mean [Fe/H] = $-0.04\pm0.27$~dex. The large scatter in metallicity suggests that a metal-rich environment may not be necessary to form massive planets at large orbital distances. We also find that the planet mass~--~host star metallicity relation for the directly imaged massive planets in wide-orbits is very similar to that found for the well studied population of short period ($\lesssim 1$~yr) super-Jupiters and brown-dwarfs around main-sequence stars.
\end{abstract}.


\section{Introduction}
\label{sec:intro}
Existing planetary search methods are constrained by severe selection effects  and detection biases \citep[e.g.][]{cum04, zak11,kip16}. However, multiple detection techniques sample different regions of the star-planet parameter space, thus providing useful insights about the rich diversity and underlying population of the planetary systems. While the transit and radial velocity methods have been successful in unraveling planet population spanning extremely close-in ($\sim$0.1~AU) to moderate orbits ($\sim$10~AU), the direct imaging is most useful for probing the planetary architecture in the outermost regions (10s-1000s~AU) of stars \citep{win15, bow16, Baron19}. The planet population discovered by the transit technique and radial velocity largely belongs to main-sequence and post-main sequence stars. In contrast, the direct imaging method has been most effective in uncovering newly formed warm and massive planets in wider orbits around nearby young stars in the solar neighborhood \citep[e.g.][]{lag14,bow16,2017AJ....154..245M, Baron19}.  

\begin{figure*}[ht]
\gridline{\fig{Planet-pops.pdf}{0.49\textwidth}{(a)}
\fig{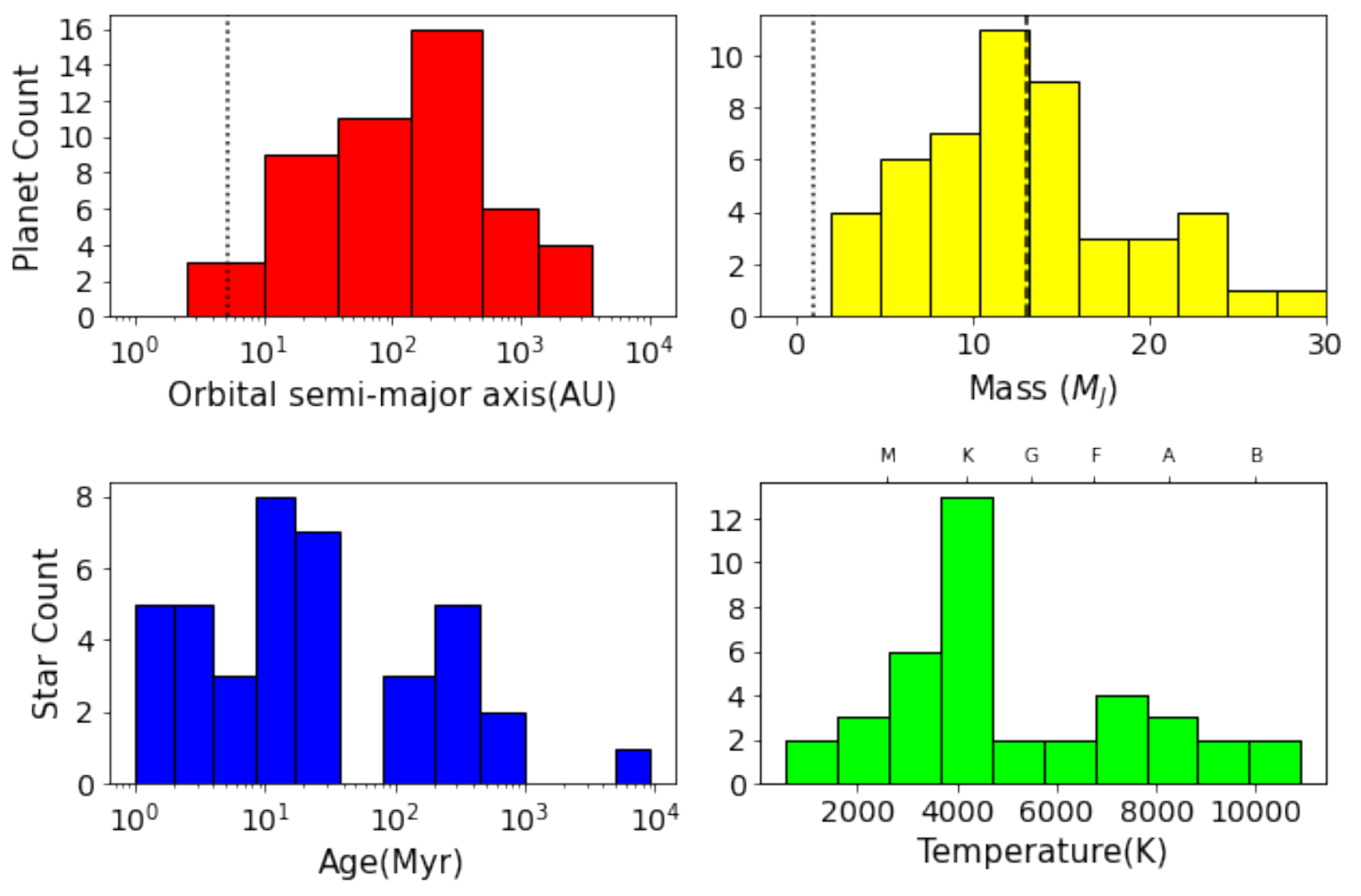}{0.52\textwidth}{(b)}}
\caption{(a) Distribution of confirmed exoplanets in mass orbital-distance plane. The orange-dots represent planets discovered by the direct imaging technique, and the blue-dots are planets discovered by other detection methods. (b) 
Histograms of orbital distance and mass of directly imaged systems (top panel) with age and temperature distribution of their stellar hosts (bottom-panel). The dotted lines in the top panel represent Jupiter's orbital distance and mass, while the dashed line at $13~M_{\mathrm{J}}$ is the minimum deuterium-burning mass limit.
\label{pops-hist}}
\end{figure*}
Following the success of the Kepler space mission, a wealth of new information has emerged about the planet population associated with main-sequence and evolved stars \citep{bor11, how12,2016AJ....152..187M,  bat14,2017AJ....154..108J,2017AJ....154..107P, nar18,  2018AJ....156..264F,  2018AJ....155...89P}. The growing number of exoplanets from space discoveries and their follow-up studies from the ground is making planetary statistics more robust and significant. Because of their large number, the statistical properties of close-in planets ($\la$~1~AU) and their host stars are relatively better studied. A great deal of research effort has been devoted to understanding the diversity of planets and the characteristics of their primary hosts. Many useful insights have been gained by studying the interdependence of planetary properties and stellar parameters \citep{1997MNRAS.285..403G,san00,san04, fis05,udr07, jon10, ghe10, Mul16,mul18, nar18, Adibekyan19}. Stellar metallicity and planet occurrence rate, for example, is one such important correlation for testing the veracity of various planet formation mechanisms under different conditions \citep[e.g.][]{udr07, mul18, san17, nar18}. However, these results have been demonstrated only for stars with close-in ($\la$~1 AU) planets that have been detected primarily by radial velocity and transit methods.

Directly imaged planets (DIP) are located at relatively large orbital distances from their host stars ($2.6-3500$~AU), which provides a unique window to probe an entirely different planetary population. While there is a general consensus that giant planets are common around high-metallicity stars compared to the low-metallicity counterparts, a clear picture is still lacking about the role of metallicity and the exact mechanism of giant planet formation at larger distances.

The majority of the 51 planetary companions discovered so far by direct imaging technique are massive planets at larger orbital distances from the host stars. 
Figure~\ref{pops-hist}(a) shows the confirmed exoplanets in a mass~-~orbital distance plane, where the segregation of planets into different populations is evident. Treating DIPs as a separate population and studying their hosts' properties can provide vital clues about the dominant mechanism of planet formation at large orbital distances from the star. The parameter space of massive planets at long orbital periods occupied by DIPs is relatively unexplored for the correlation studies of host star-planet properties. Also, the high-mass limit of wide-orbit planets overlaps with the low-mass tail of brown-dwarfs and sub-stellar companions. Therefore, in certain cases, the limitation of low-number DIP statistics can be partly overcome by a complementary study of known brown-dwarf companions sharing the same parameter space \citep{ma14,vig17, nie19}.  Therefore, it is essential to investigate the role, if any, of the host-star metallicity in influencing the process of the giant planet and brown-dwarf formation over a wide range of astrophysical conditions.

We have examined the confirmed list of DIP hosted on the NASA's Exoplanet archive \citep{2013PASP..125..989A}\footnote{https://exoplanetarchive.ipac.caltech.edu/index.html}. 
The available stellar and planetary parameters are compiled from the composite planet data table for known exoplanets and published literature. Each of these systems has been studied and discussed in depth by individual discovery and follow-up papers. However, there are limited instances where the DIP distribution and stellar properties are studied as separate ensemble \citep{neu12, bow16}.  

Out of the 45 stars hosting DIPs listed in the Table~\ref{star_params} taken form the NASA Exoplanet archive, we could cross-match 42 of them with the GAIA DR2 catalog of which  $T_\mathrm{eff}$ and luminosity was available for 26 stars (for cross-matching see  \cite{2020AJ....159..194V}). The atmospheric properties of the stars hosting these wide orbit companions are not very well studied, and most notably, the metallicity is known only for 14 such systems. 

In general, previous studies \citep{buc14,san17,nar18,2018ApJ...853...37S} have shown that the average metallicity of the host star increases as a function of planetary mass. However, the trend reverses for most planetary-mass above 4-5 $M_{J}$ \citep{nar18,san17,2018ApJ...853...37S,Mal19}. These results suggest the possibility of two planet formation scenarios with the Jupiter-like planets ($0.3-5M_{J}$) likely formed by the core-accretion process\citep[e.g.][]{MIZ80,Pollack96,Ida04,2012A&A...541A..97M} and the massive super-Jupiters ($> 5M_{J}$) via the disk instability mechanism \citep[e.g.][]{Boss1836, 2002Sci...298.1756M, Boss02, 2007ApJ...662.1282M, san17, nar18, god19}. These findings, backed by large statistics, truly reflect the underlying metallicity-mass distribution of compact planetary systems (orbital period $\la$~ 1~yr). This raises another important question whether or not such trends hold for planets formed at vast orbital distances from the central star. Since DIPs are found at large distances from their host stars, this planet-population motivates us to explore the mass-metallicity relationship for giant planet populations at large distances in light of various planet formation scenarios. This paper has used high-resolution spectra available from various public archives to determine the stellar parameters and metallicity of 18 stars hosting DIPs in a consistent and homogeneous way to study the various correlation among stellar and planetary properties.

The rest of the paper is organized as follows. In section 2, we give a brief overview of directly imaged planetary systems. We describe our sample and give the selection criteria in section 3. Our methodology and Bayesian approach used for the estimation of various stellar parameters is discussed in section 4. In section 5, we discuss our results and compare them with previous findings. Finally, we give our summary and conclusions in section 6.
 

\begin{figure}[t]
\includegraphics[width=3.5in]{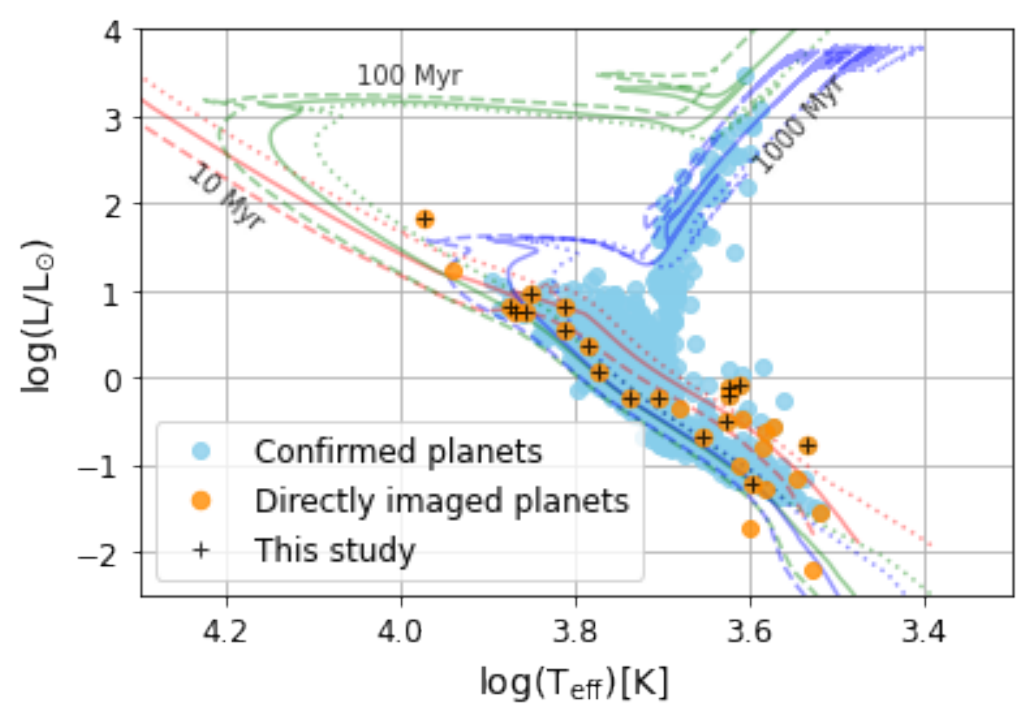}
\caption{The location of confirmed exoplanet hosting stars in the HR diagram. The $T_{\mathrm{eff}}$ and luminosity $L$ are compiled for 2831 confirmed planet hosts that are cross-matched with $Gaia$ DR2 catalog. The sky-blue circular symbols represent host stars of planets discovered by indirect methods. The orange circles show the stellar companions of directly imaged planets. A subset of 18 DIP host stars used in the present study is indicated by orange circles with `+' symbol in the middle. Isochrones computed using \cite{2016ApJ...823..102C} are shown for three age groups (red-line: 10 Myr, green-line: 100 Myr and blue-line: 1000 Myr) and metallicity range: solid-line [Fe/H]=0~dex, dotted-line [Fe/H]=0.5~dex and dashed-line [Fe/H]=-0.6~dex.}
\label{hr_dia}
\end{figure}


\section{Directly Imaged Systems}
Of the $4200+$ confirmed planets, direct imaging technique accounts for the discovery of 51 planetary-mass objects around 45 stars. Among these, 40 are in a single planetary system, and four are in multi-planetary systems -LkCa 15,  TYC 8998-760-1, and PDS70 with two planets each and HR8799 with four. The majority of them are discovered from deep imaging surveys of nearby star forming regions. These  planet search programs largely target  young pre-main sequence stars that belong to nearby stellar associations and moving groups, all within 200~pc of the Sun \citep{bow16}. The high luminosity of planets at early formation stage make them amenable for the direct imaging. Further, the high-resolution and high-contrast imaging of planets is facilitated by the adaptive optics technology and stellar coronagraphy. With advanced differential imaging and psf extraction techniques, new generation of instruments, e.g. Gemini Planet Imager (GPI), ScExAO on Subaru and SPHERE on VLT, are capable of probing Jupiter-mass planetary companions within a few mas separation from the central star. 
  
Masses of self-luminous planets are inferred from hot-star evolutionary tracks and infrared fluxes, but in some cases, they are well constrained by precise astrometric measurements \citep{bar03, wan18, sne18, nie19, wag19}. The onset of  deuterium burning limit  ($\sim 13 M_{\mathrm{J}}$) is  a commonly used criteria to separate a planet from a brown-dwarf \citep{bur97,sau08,spi11}. However, by taking different composition and formation scenarios into account, the upper cut-off range could be as high as  $25-30~M_{\mathrm{J}}$  \citep{bar10, sch11}. We acknowledge this ambiguity of overlapping mass range, but we clump all directly imaged objects up to $\sim 30 M_{\mathrm{J}}$ in the  DIP  category for the present work. 

The histogram shown in Figure~\ref{pops-hist}(b) reveals that except for one case\footnote{CFBDSIR J145829+101343b is the closest planet at orbital distance 2.6~AU from the central star that is resolved by the direct imaging.}, the projected semi-major axis distance of all DIPs are larger than the Jupiter's orbital distance. The distribution peaks at an orbital distance of 150-500~AU  and extend up to $\approx 3500$~AU. The lower limit of the distribution is set by the inner working angle of the coronagraph, while the drop beyond few thousand AU is influenced by the limited sensitivity to detect the positional change of planets in long-period orbits. 

The median mass of the DIP population is about $\approx 12.5\,M_{\mathrm{J}}$ with lowest mass object $2\,M_{\mathrm{J}}$ and about half the number more massive than $13\,M_{\mathrm{J}}$. Most stellar hosts of these planets are also relatively young, i.e., $\approx 75\%$  below the age of $\sim 100$~Myr and more than two-thirds of the total belonging to the late spectral types with $T_{\mathrm{eff}} \leqslant 4500$~K. From the literature, we also find evidence of circumstellar disk around 22  such systems. 

The equilibrium temperature of imaged planets ranges from $300-2800$~K, though most of them are above $1600$~K. The projected angular separation between the host-star and planet varies by four orders of magnitude ranging from $\approx10^{-2}-10^{2}$ arc-sec. A large angular separation from the central star and inherent brightness due to their high temperature make this giant planet population ideal for direct detection \citep{tra10}. 

We note that the current DIP sample is not a true representative of the underlying population of planets in outer orbits. It is heavily biased towards young, hot, more-massive ($\geqslant 4M_{\mathrm{J}}$) companions of young stars. The complexity of high-contrast instruments and the limitation of observing a single object at a time also makes the discovery rate slow. Studying  DIP hosts spectroscopically is a major challenge because of their wide spectral range and complexities (veiling, extinction, etc) associated with young and pre-main sequence stars. Therefore, it is also difficult to apply a strictly uniform and homogeneous methodology for the whole sample's characterization.

\section{Sample Selection} 
The NASA Exoplanet archive has 3185 stars with confirmed planets found by various discovery methods. We found 2831 out of  stars cross-matched with the GAIA DR2 catalog, which has the most accurate parallaxes and precise multi-band photometry of all-sky stellar sources down to magnitude $G\approx 21$. Figure~\ref{hr_dia} shows the location of these stars in the HR diagram with $T_{\mathrm{eff}}$, and stellar luminosity is taken from the GAIA catalog.

The archive also contains the list of 45 host stars of directly imaged planets given in Table~\ref{star_params}. Of these, 42 are found in the GAIA DR2 catalog, and their position in the HR diagram is also shown in Figure~\ref{hr_dia}.
The summary of astrophysical parameters of the DIP host stars listed in Table~\ref{star_params} and our selection criteria for spectroscopic analysis is as follows: 

\begin{itemize}
    \item We searched various public archives for the availability of high-resolution optical spectra for individual DIP hosts and also surveyed the literature on their metallicity. Based on these findings, we separated the 45 DIP host stars in Table~\ref{star_params} into three distinct groups demarcated by horizontal lines. 
    
    \item The first 18 stars in Table~\ref{star_params} is a subsample of DIP host stars analyzed in this paper for which the spectra are available from public archives, but literature metallicity is known only for ten targets. These stars have an effective temperature range between 4059-10690~K and G-band magnitude smaller than $\sim13$. For this subsample, we determined the atmospheric parameters and metallicity [Fe/H]  homogeneously for the first time. We obtained high-resolution, high-SNR spectra for 14 targets from the ESO science archive facility\footnote{http://archive.eso.org/scienceportal/}  and for four targets from Keck\footnote{https://koa.ipac.caltech.edu/cgi-bin/KOA/nph-KOAlogin} archive. The ESO's Science Portal provides access to the already reduced and wavelength calibrated data.  Details of original spectra, e.g., telescope/instrument,  resolution, wavelength coverage, and SNR, are listed in Table~\ref{archival_spectra}.

    \item In the 2nd group of Table~\ref{star_params}, there are 4 DIP host stars for which the metallicity is taken from the literature. The last 23 DIP hosts belonging to the 3rd group in Table~\ref{star_params} are not analyzed in this paper because majority of them are fainter  ($m_v>13$). For these stars  either the spectra was not available in the public domain or the quality of the data was poor (low-SNR). This group also includes some of the hot and very rapidly rotating stars ($v\cdot\sin i> 160$~Km/s), which do not have clear spectral features and reliable atmospheric models for parameter estimation.   
    
    \item Most stellar parameters listed in Table~\ref{star_params} are taken from the NASA Exoplanet Archive. Furthermore, we cross-checked the accuracy of these parameters and replaced the missing values with those from the discovery and relevant follow up papers. The log$\,g$ values marked by `*' symbols are not listed in the standard archives(such as Nasa exoplanet archive), and we have calculated them from stellar mass and radius values available from the literature.  \end{itemize}

\section{Estimation of stellar parameters}
Spectral synthesis and equivalent width (EW) method are two commonly used techniques to derive the stellar parameter of interest from a high-resolution spectra of stars \citep{1994AJ....107..742G,2002A&A...383..227E, nis18, bla19, jof19}. Despite intrinsic differences, each method requires the proper prescription of a stellar atmospheric model, a well-characterized atomic line list, reference solar abundance, and the radiative transfer code. Most notably, the relevant model parameters in both methods are allowed to vary, and a least-squares minimization is performed to reach the convergence. For example, in the EW case, the desired parameters are those for which the correlation between abundances and equivalent widths (excitation equilibrium and ionization balance) is minimized to zero. In spectral synthesis, theoretical spectra are iteratively generated from the model atmosphere and compared with the observed spectra of the star until a best match is found. The parameters of the best-matched spectra are the closest that describe the properties of the real star. The spectral synthesis method, which we adopted for our Bayesian model, is also suitable for analyzing young and fast-rotating stars present in our sample. 

\subsection{Generation of model spectra}
We adopted the Bayesian approach to infer the stellar parameters from the model spectra generated using iSpec --an integrated open-source software \citep{bla14}. iSpec is a python wrapper that incorporates various radiative transfer codes, stellar atmospheric models, and many ready-to-use tools to derive stellar parameters and abundances \citep{bla14, Blac14}. As explained in the next section, we use iSpec only as a back-end module to generate synthetic spectrum on-the-fly to navigate the stellar parameter space for determining the posterior distribution of T$_\mathrm{eff}$, log$g$, [Fe/H] and $v\cdot \sin i$ for our 18 target stars. For generating the model spectra in iSpec, we selected the radiative transfer code SPECTRUM (R.O Grey) because of its faster performance compared to other codes \citep{Blac14}. This code assumes the local thermodynamic equilibrium condition and requires a grid of plane-parallel model atmosphere as input. We chose ATLAS9 model atmosphere that has grid sampling of 250~K in $T_\mathrm{eff}$, 0.5~dex in $\log\,g$ and the metallicity sampled over  0.4,0.2,0.0,-0.5,1,-1.5,-2,-2.5,-3,-4 grid points \citep{2003IAUS..210P.A20C}. To generate model spectra for intermediate values, iSpec uses interpolation. The solar abundances are taken from \cite{2009ARA&A..47..481A} and the atomic linelist from the VALD database \cite{pis95} that also comes bundled with iSpec. We also adjusted the oscillator strengths and broadening parameters for some of the lines in our line list to improve our ability to model the stellar spectrum in the 600-620~nm wavelength regions, following the procedure given by \citep{2007MNRAS.379..773S}. The microturbulence and macro turbulence velocities were internally calculated by iSpec using empirical relations \citep{Blac14}.\\ 
\subsection{Data Preparation}
Doing Bayesian analysis on the whole spectrum is computationally prohibitive. To reduce the computational load, we considered three distinct wavelength regions of the spectrum. These regions are free from telluric lines and also serve good proxies for different stellar parameters without any degeneracy \citep{pet18}. 

\centerwidetable
\startlongtable
\begin{longrotatetable}
\begin{deluxetable*}{lccccccccccc}
\tabletypesize{\scriptsize}
\tablecaption{Stellar parameters of directly imaged planet host stars.} 
\label{star_params}
\tablehead{
\\
  &  \multicolumn{6}{c}{\bf{Literature values}} & 
     & \multicolumn{4}{c}{\bf{This paper}}
    \\
 \cline{2-7}    \cline{9-12} \vspace{-3mm}
\\ 
\colhead{\bf{Star}} \vspace{-1mm} & \colhead{Age} &  \colhead{$M_{P}$} & \colhead{$T_{\mathrm{eff}}$} & \colhead{log$\,g$}  & 
\colhead{[Fe/H]} & \colhead{$v\cdot \sin i$} & \colhead{Ref} & \colhead{$T_{\mathrm{eff}}$} & \colhead{log$\,g$}  &
\colhead{[Fe/H]} & \colhead{$v\cdot \sin i$} \vspace{-3mm} \\
\colhead{} &\colhead{[Myrs]} &\colhead{[$M_{\mathrm{J}}$]} & \colhead{[K]} & \colhead{[cms$^{-2}$]} &\colhead{[dex]} & \colhead{[kms$^{-1}$]} & \colhead{} & \colhead{[K]} & \colhead{[cms$^{-2}$]} & \colhead{[dex]} & \colhead{[kms$^{-1}$]}\\   \vspace{-2mm}
}
\startdata
\multicolumn{12}{c}{DIP host stars analyzed in this paper}
\\[2ex]
\hline
HD 106906 & 13 & $11\pm2$& $6516^{+165}_{-165}$ & -- & -- & $55^{+4.0}_{-4.0}$& 1 2 A & $6798^{+20}_{-40}$ & $4.23^{+0.02}_{-0.05}$ & $0.04^{+0.01}_{-0.02}$ & $49.12^{+0.22}_{-0.17}$\\
AB PIC & 17.5 & $13.5\pm 0.5$ & $5378^{+55}_{-55}$& $4.44\pm{0.21}$ & $-0.05\pm0.04$ & $11.5^{+0.1}_{-0.1}$ & 3 4 5 6 A&$5285^{+10}_{-9}$  & $4.53^{+0.01}_{-0.01}$ & $0.04 \pm 0.02$ & $10.35^{+0.06}_{-0.04}$  \\
GJ 504 & 160 & $4.0^{+4.5}_{-1.0}$ & $6234^{+25}_{-25}$ & $4.33\pm0.10$&$0.28\pm0.03$& $7.4^{+0.5}_{-0.5}$ & 7 8 B& $6291^{+14}_{-16}$ & $4.34^{+0.01}_{-0.02}$ & $0.27^{+0.02}_{-0.03}$ & $5.47^{+0.12}_{-0.17}$ \\
HN Peg & 237 &$ 21.99\pm9.43$ & $6034$ & $4.48$ & $-0.02\pm0.02$ & $10.6^{+0.5}_{-0.5}$& 9 10 11 A & $6186^{+14}_{-7}$ & $4.48^{+0.03}_{-0.02}$ & $0.00^{+0.01}_{-0.02}$ & $8.73^{+0.06}_{-0.05}$  	\\
51 Eri & 20 & 2.0 & 7146  & $3.99\pm0.24$ & $0.24^{+0.35}_{-0.35}$ & $71.8^{+3.6}_{-3.6}$  & 12 13 A& $7276^{+11}_{-9}$  & $4.08^{+0.03}_{-0.02}$ & $0.13^{+0.03}_{-0.02}$ & $65.19^{+0.14}_{-0.17}$\\
HR 2562& 600 & $30\pm15 $ & 6534 & $4.18^{+0.04}_{-0.05}$ & $0.08$ & 14 15 16 & 16 17 18 C& $6785^{+29}_{-27}$ & $4.40^{+0.04}_{-0.05}$ & $0.21^{+0.02}_{-0.03}$ & $43.51^{+0.15}_{-0.17}$  \\
Fomalhaut & 440 & $2.6\pm0.9$ & 8689 & $4.11^{+0.85}_{-0.85}$ & $0.27^{+0.19}_{-0.19}$ & $91.06^{+0.5}_{-0.5}$ &  17 18A & $8508^{+43}_{-12}$ & $4.02^{+0.01}_{-0.02}$ & $0.13^{+0.05}_{-0.03}$ & $75.31^{+0.57}_{-0.14}$ \\
HR 8799 & 30 & $7^{+4}_{-2}$ &$7376^{+218}_{-217}$ & $4.22$ & $-0.5$ & --  & 19 20 21 22 A&$7339^{+4}_{-5}$ & $4.19^{+0.01}_{-0.02}$ & $-0.65^{+0.02}_{-0.01}$ & $34.79^{+0.24}_{-0.13}$\\ 
& & $10^{+3}_{-3}$ &&&&&&&& \\
& & $10^{+7}_{-4}$ &&&&&&&& \\
HD 203030 & 220 & $24.09^{+8.38}_{-11.52}$ & $5472^{+84}_{-61}$ & $4.50^{+0.05}_{-0.05}$ & $0.06^{+0.07}_{-0.07}$ & $6.3^{+0.3}_{-0.3}$& 15 21 23 24 D &  $5603^{+10}_{-8}$ & $4.64^{+0.03}_{-0.01}$ & $0.30^{+0.02}_{-0.01}$ & $5.62^{+0.13}_{-0.14}$ \\ 
HD 95086 & 17 & $5\pm2$ & $7750\pm250$ & $4.0\pm0.5$ & $-0.25\pm0.5$ & $20\pm10$ & 25 26 27 A&$7883^{+43}_{-65}$  & $4.58^{+0.02}_{-0.06}$ & $0.14^{+0.05}_{-0.04}$ & $17.14^{+1.31}_{-0.52}$ \\
bet Pic & 12.5 & $11\pm2$ & $8052$ & 4.15 & $-0.1\pm0.2$& 122.0 & 3 28 29 A&$7890^{+13}_{-17}$ & $3.83^{+0.03}_{-0.02}$ & $-0.21^{+0.03}_{-0.02}$ & $116.41^{+2.32}_{-2.05}$  \\
HIP 78530 & 11 & $23\pm1$ &10500 & --& -- & -- & 30 31 A&$ 10690^{+24}_{-10}$ & $4.68^{+0.02}_{-0.01}$ & $-0.50^{+0.03}_{-0.01}$ & $144.74^{+0.55}_{-3.14}$ \\
LkCa 15 & 1 &--& $4210^{+185}_{-199}$ & -- & -- & $13.90^{+1.20}_{-1.20}$ & 23 32 33 D&$4589^{+7}_{-7}$ & $3.65^{+0.01}_{-0.01}$ & $0.26^{+0.01}_{-0.01}$ & $16.82^{+0.1}_{-0.2}$ \\ 
PDS70 & 5 &$8\pm6$ &  $4225^{+242}_{-71}$ & -- & -- & --&23 34 35 36 A &  $4152^{+5}_{-9}$ & $3.68^{+0.01}_{-0.01}$ & $-0.11^{+0.01}_{-0.01}$ & $17.27^{+0.1}_{-0.1}$ \\
&& $8\pm4$ &&&&&&&& \\
CT CHA & 2 & $17\pm6$ & $4200^{+211}_{-115}$ & -- & -- & $12.80^{+1.7}_{-1.7}$  & 8 23 37 38 B & $4403^{+6}_{-10}$ & $3.66^{+0.01}_{-0.01}$ & $-0.56^{+0.01}_{-0.01}$ & $13.97^{+0.10}_{-0.15}$\\
GQ LUP & 1 & $20$& $4092^{+211}_{-165}$ & -- & -- & --  &23 39A &$4416^{+3}_{-5}$ & $3.65^{+0.01}_{-0.04}$ & $-0.35^{+0.01}_{-0.01}$ & $6.33^{+0.03}_{-0.07}$ \\ 
ROXs 12 & 6 &  $16\pm4$ & $3850^{+100}_{-70}$ & -- & -- & -- & 40 41 D& $4059^{+3}_{-4}$ & $3.71^{+0.01}_{-0.01}$ & $0.14^{+0.01}_{-0.01}$ & $7.20^{+0.03}_{-0.04}$  \\ 
GSC 06214-00210    & 11 & $16\pm1$ & $4200^{+150}_{-150}$ & -- & -- & --& 3 4 D & $4119^{+6}_{-13}$ & $3.70^{+0.01}_{-0.04}$ & $-0.06^{+0.01}_{-0.01}$ & $4.24^{+0.04}_{-0.05}$  \\[2ex]
\hline
\multicolumn{12}{c}{DIP host stars  with metalicity from literature}
\\[2ex]
\hline
HIP 65426 & 14 & $9.0\pm3.0$ & $8840^{+200}_{-200}$ & $5.00^{+0.13}_{-0.18}$ & $-0.03^{+0.10}_{-0.10}$ & $299\pm9$& 42 43 44 &--&--& --& --\\ 
Kap And & 220 & $13.616^{+23.04}_{-1.05}$ & $10900^{+300}_{-300}$ & $3.78^{+0.08}_{-0.08}$ & $-0.36^{+0.09}_{-0.09}$ & $176$&  45 46 47& -- & --&--& -\\
GU PSC & 100 & $11.3\pm1.7$ &$3250\pm32$ & $4.75^{+0.07}_{-0.07}$ & $0.10^{+0.13}_{-0.13}$ & $23.0^{+0.14}_{-0.14}$ & 48 49& -- &-- & --&-- \\ 
Ross 458 & 475 & 6.28536 & $3600\pm73$ & $4.71^{+0.08}_{-0.05}$ & $0.25^{+0.08}_{-0.08}$ & $9.75$ &50 51 52 53& --	&-- &-- & --\\[2ex]
\hline
\multicolumn{12}{c}{ DIP host stars that are not analyzed in this paper}
\\[2ex]
\hline
ROXs 42B & 2 & $9\pm3$ & $3850^{+199}_{-394}$ & -- & -- & -- & 23 54 55&  &  &  &  \\ 
2MASS J02192210-3925225 & 20 &$13.9\pm1.1$&$3064^{+76}_{-76}$ & $4.59^{+0.06}_{-0.06}$ & -- & $6.5^{+0.04}_{-0.04}$& 56 & -- &--&--&--\\
NAME Oph 11 & 2 &$14^{+6}_{-5}$ &$2375^{+175}_{-175}$& $4.25^{+0.50}_{-0.5}$ &--& --& 55 57 & -- &-- &--& --\\
VHS J125601.92-125723.9& 225 &$11.2^{+9.7}_{-1.8}$&$2620^{+140}_{-140}$ & $5.05^{+0.10}_{-0.10}$ & -- & --&58 & -- &-- & -- &-- \\
WISEPJ121756.91+162640.2A & 6000 &$22\pm2$&$575\pm25$ & $5.0\pm0.1$ & -- & -- & 59&-- & -- & -- &  -- \\
CFBDSIR J145829+101343 & -- & $10.5^{+4.5}_{-4.5}$& $581^{+24.5}_{-24.5}$ & $4.73^{+0.28}_{-0.28}$ & -- & --& 60 & --  &-- & -- & -- \\
1RXS J160929.1-210524   & 11 &$8\pm1$ & $4060^{+300}_{-200}$ & $4.19^{+0.09}_{-0.06}*$ & -- & -- &3 4 23& -- &-- & -- & -- \\
2MASS J21402931+1625183 A & -- &$20.95^{+83.79}_{-20.95}$&
$2300^{+80}_{-80}$ & $5.17^{+0.24}_{-0.61}*$ & -- & --& 23 61 & -- & -- & --& --\\
2MASS J22362452+4751425   & 30 &$12.5^{+1.5}_{-1.5}$& $4045^{+35}_{-35}$ & $4.60^{+0.04}_{-0.04}*$ & -- & -- &5 23 62& -- &-- & -- & -- \\
DH Tau & 1 & $11^{+10}_{-3}$ & $3751^{+501}_{-148}$ & $5.06^{+0.25}_{-0.20}*$ & -- & --& 23 33 63 & -- &-- & -- & -- \\
TYC8998-760-1  & 17 & $14^{+3}_{-3}$ & $4753^{+10}_{-10}$ & $4.44^{+0.01}_{-0.01}*$ & -- & -- &23 64& --  &-- & -- & -- \\
& & $6^{+1}_{-1}$ & & & & & & & & &  \\
USco1556 A & 7.5 &$15^{+2}_{-2}$ & $3400^{+100}_{-100}$ & $4.49^{+0.08}_{-0.10}*$ & -- &-- &23 65 66& -- &-- &--& --\\
USco1621 A & 7.5 & $16^{+2}_{-2}$ &$3460^{+100}_{-100}$ & $4.25^{+0.09}_{-0.11}*$ & -- & -- &23 65 66& --&-- & -- & --\\
HIP 79098 AB & 10 & $20.5\pm4.5$ &$9193^{+61}_{-93}$ & -- & -- & --&23 67& -- & --& -- \\
2MASS J01225093-2439505 & 120 & $24.5^{+2.5}_{-2.5}$ & $3530^{+50}_{-50}$ & -- & -- &$14.2^{+3.2}_{-3.2}$ & 68 & -- &-- &-- & -- \\
NAME SR 12 AB    & 2.1 &$13^{+7}_{-7}$ & $3828^{+516}_{-379}$ & -- & -- & --  & 23 69 & -- &-- & -- & --\\
USco CTIO 108 & 11 &$14^{+2}_{-8}$ & $2700^{+100}_{-100}$ &-- & -- & -- & 4 70 & -- &-- & -- & -- \\
WD 0806-661 & 1500 & $7.5^{+1.5}_{-1.5}$ & $9552^{+54}_{-1931}$ & -- & -- & -- &  23 71 72 &-- & -- & -- & -- \\
FU Tau & 1 & 16 & $2838$ & -- & --& --& 33 73 & -- & --&  -- & --  \\
2MASS J04414489+2301513 & 1 & $7.5^{+2.5}_{-2.5}$ &-- & -- & -- &-- &  74& -- & -- & -- & -- \\
2MASS J12073346-3932539  & 8 & $4^{+1}_{-1}$ &-- & -- & --& --& 75 76& --& -- & -- & -- \\
CHXR 73  & 2 &$12.57^{+8.38}_{-5.24}$& -- & -- & -- &--& 38 77 & -- & -- & -- & -- \\
2MASS J01033563-5515561 & 30 &$13\pm1$& -- & -- & -- &--& 38 78 & -- & -- & -- & -- \\
\enddata
\tablecomments{1.\cite{2011ApJ...738..122C}; 
2.\cite{2014ApJ...780L...4B};
3.\cite{Ujjwal_2020};
4.\cite{tep1};
5.\cite{ghe10};
6.\cite{2006Torres};
7.\cite{2013ApJ...774...11K};
8.\cite{2015Mal};
9.\cite{2007ApJ...654..570L};
10.\cite{2009Ram};
11.\cite{2015A...573A..17B};
12.\cite{2015Sci...350...64M};
13.\cite{2017AJ....153...21L};
14.\cite{2016ApJ...829L...4K}; 
15.\cite{mal12};
16.\cite{2003AJ....125.1598L};
17.\cite{2008Sci...322.1345K};
18.\cite{2012ApJ...754L..20M};
19.\cite{2008Sci...322.1348M};
20.\cite{GC};
21.\cite{GAIA};
22.\cite{2010Natur.468.1080M};
23.\cite{2006ApJ...651.1166M};
24.\cite{Faherty_2009};
25.\cite{2013ApJ...779L..26R};
26.\cite{2013ApJ...775L..51M};
27.\cite{2013ApJ...775L..40M};
28.\cite{2018NatAs...2..883S}; 
29.\cite{1994AAS...185.4812H};
30.\cite{2015ApJ...802...61L};
31.\cite{2012ApJ...746..154P};
32.\cite{2012ApJ...745..119N};
33.\cite{1995ApJS..101..117K};
34.\cite{KB};
35.\cite{2019NatAs...3..749H};
36.\cite{ refId0}
37.\cite{SN};
38.\cite{2011ApJS..193...11M};
39.\cite{2008Neuh};
40.\cite{2014ApJ...781...20K};
41.\cite{2017AJ....154..165B};
42.\cite{CD};
43.\cite{2018AJ....155..149B};
44.\cite{2017A...605L...9C};
45.\cite{RZ};
46.\cite{BC};
47.\cite{2013ApJ...779..153H};
48.\cite{2014ApJ...787....5N};
49.\cite{2014ApJ...788...81M,2014ApJ...792...37M};
50.\cite{2011ApJ...732L..29R};
51.\cite{2014ApJ...791...54G};
52.\cite{2010ApJ...725.1405B};
53.\cite{2010MNRAS.407.1657H};
54.\cite{2014ApJ...787..104C};
55.\cite{2005AJ....130.1733W};
56.\cite{2015ApJ...806..254A};
57.\cite{2007ApJ...660.1492C};
58.\cite{2015ApJ...804...96G};
59.\cite{2014ApJ...780...62L};
60.\cite{2011ApJ...740..108L};
61.\cite{2010ApJ...711.1087K};
62.\cite{2017AJ....153...18B};
63.\cite{PK};
64.\cite{2020MNRAS.492..431B};
65.\cite{CP};
66.\cite{2019AJ....158..138S};
67.\cite{JM};
68.\cite{2013ApJ...774...55B};
69.\cite{2011AJ....141..119K};
70.\cite{2008ApJ...673L.185B};
71.\cite{2012ApJ...744..135L};
72.\cite{2011ApJ...730L...9L};
73.\cite{2009ApJ...691.1265L};
74.\cite{2010ApJ...714L..84T};
75.\cite{DT};
76.\cite{2007ApJ...657.1064M};
77.\cite{2006ApJ...649..894L};
78.\cite{Del13};\\ The instrument listed in the reference column are as follows --
A: HARPS; B :UVES ; C : FEROS ; D :HIRES 
 }
\end{deluxetable*}
\end{longrotatetable}

\begin{deluxetable}{cccccc}[t]
\tablecaption{Properties of archival spectra and instrument used. \bf{The last column refers to the median SNR of all DIP host stars spectra observed with each instrument.} \label{archival_spectra}}
\tablewidth{0pt}
\centering
\tablehead{
\colhead{Instrument} &\colhead{Spectral Range} & \colhead{Resolution} &\colhead{SNR}
\\
\colhead{} &\colhead{in nm} & \colhead{} & \colhead{} & \colhead{}  
}
 \startdata
HARPS & 378.2-691.3 & 115000 & 174\\
UVES & 472.7-683.5 & 74450 & 218\\ 
FEROS & 352.8-921.7 & 48000 & 305\\
HIRES & 336.0-810.0 & 85000 & 60\\
\enddata

\end{deluxetable}
The first region is the Mg-I triplet (5150-5200 \AA), which is sensitive to $\log\,g$. The second region (6000-6200~\AA) includes a significant number of well-isolated and unresolved spectral lines that are sensitive to $v\cdot\sin i$ and [Fe/H], and the third region (6540-6590~\AA) covers the $H_{\alpha}$ line whose outer wings are sensitive to $T_{\mathrm{eff}}$. We have used all the three regions for most targets except for HIP~78530, which show severe line blending due to fast rotation. In that case, we have used only Mg-I triplet and $H_{\alpha}$ segments.

\begin{figure}[t]
\includegraphics[width=3.5in]{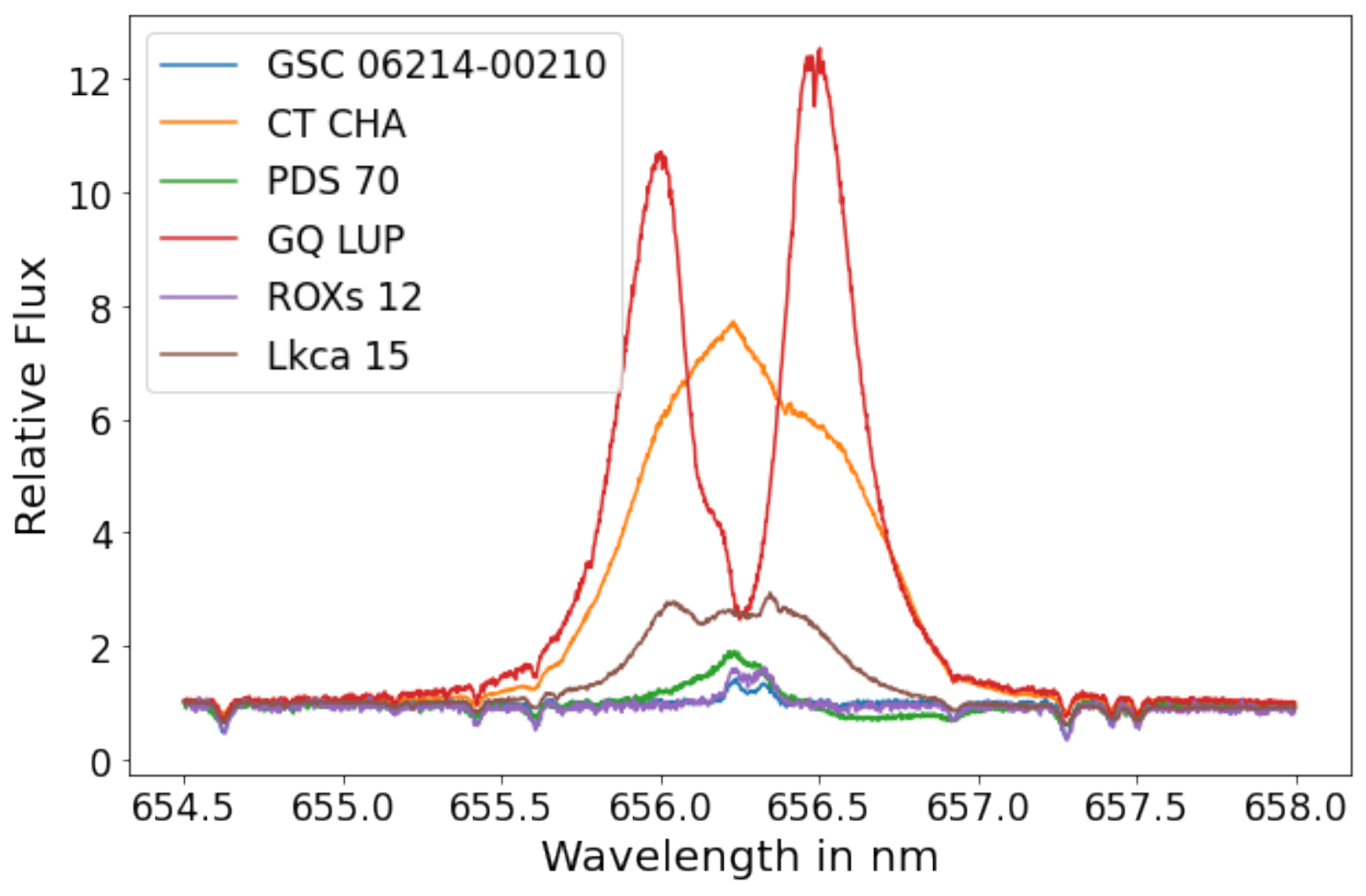}
\caption{The presence of veiling inferred from H$_\alpha$ emission line seen in the spectra of stars 13-18 in Table~\ref{star_params}.}
\label{veiling_spectra}
\end{figure}

Additionally, some of the stars in our sample (S.No 13-18 in Table~\ref{star_params}) have emission features that indicate the presence of an accretion disk around the star. The characteristic veiling dominated $H_{\alpha}$ emission for these stars is shown in Figure~\ref{veiling_spectra}. This accretion shocked region on the stellar surface generates the veiling continuum and decreases the depth of the stellar absorption lines \citep{1998ApJ...509..802C}.  Since we don't have reliable models for emission lines (such as the $H_{\alpha}$), we chose a less contaminated and emission-free regions  5900-5965{\AA} for deducing the stellar parameters \citep{2002A&A...391..595S,2003A&A...408..693S}. In addition, we included  6100-6200{\AA} segment for Lkca15, Ross12, PDS~70, and  GSC 06214-00210   together with  5900-5965{\AA} for determining stellar parameters since this region also lacks emission lines. In the Bayesian analysis discussed in the next section, we considered veiling as a free parameter to account for the excessive line filling due to the accretion, following the procedure by \cite{2002A&A...391..595S,2003A&A...408..693S}.    

Individual spectra of stars come from single-object spectroscopic observations from the different instruments. The FITS files contain a 1D spectrum with the specification of wavelength, flux, and flux-errors. If the flux-error was not specified, we assumed the errors to be limited by the photon-noise. A certain amount of pre-processing was needed to prepare the data for further analysis. We used standard packages in IRAF \footnote{IRAF is distributed by the National Optical Astronomy Observatories, which is operated by the Association of Universities for Research in Astronomy, Inc., under contract to the National Science Foundation.} for continuum normalization and the radial velocity correction in the spectra. The model spectrum was generated at the same wavelength grid as the observed spectrum. 

\subsection{Bayesian Inference and MCMC Sampler} 
\label{Bayesian}

We chose the Bayesian  approach  for probabilistic inference because it eliminates the dependence of derived stellar parameters on the initial guess values and also places realistic constraints on the errors \citep{shk07}. We denote our minimal set of model parameters as $\theta~\equiv~\{T_\mathrm{eff},\log g, \mathrm{[Fe/H]}, v.\sin i\}$ and observed stellar spectrum as $D~\equiv~\{y_{\mathrm{data}},y_{\mathrm{err}},\lambda\}$, where $y_{\mathrm{data}}$ is the measured flux at wavelength $\lambda$ and associated  uncertainty $y_{\mathrm{err}}$. The model predicted normalized flux $y_{\mathrm{mod}}(\theta,\lambda)$ is calculated from first principles using radiative transfer code and appropriate model of stellar atmosphere. The goal is to find  \emph{posterior} $p(\theta|D)$ which is the most likely distribution  of the model parameters $\theta$ conditioned on the observed data $D$. We know, from  Bayes's theorem 
\begin{equation}
 p(\theta|D)=\frac{p(\theta)p(D|\theta)}{p(D)}
 \label{eq1}
\end{equation} 

where $p(D|\theta)$ is the \emph{likelihood} of observing spectra $D$, given the set of model parameters $\theta$, $p(\theta)$  is \emph{prior function}. The term $p(D)$ in the denominator of eq.~\ref{eq1}  is a normalization constant, also called \emph{evidence}, which is hard to compute, but not required when we use a sampler. Note that each term in eq.~\ref{eq1} is a probability density function whose analytical form is rarely known in practice. The Markov-Chain Monte-Carlo (MCMC) process allows us to numerically estimate the parameters by randomly drawing a sequence of samples from the posterior distribution of model parameters constrained by the data \citep{hog18}.   
\begin{figure}[t!]
\includegraphics[width=1.0\linewidth]{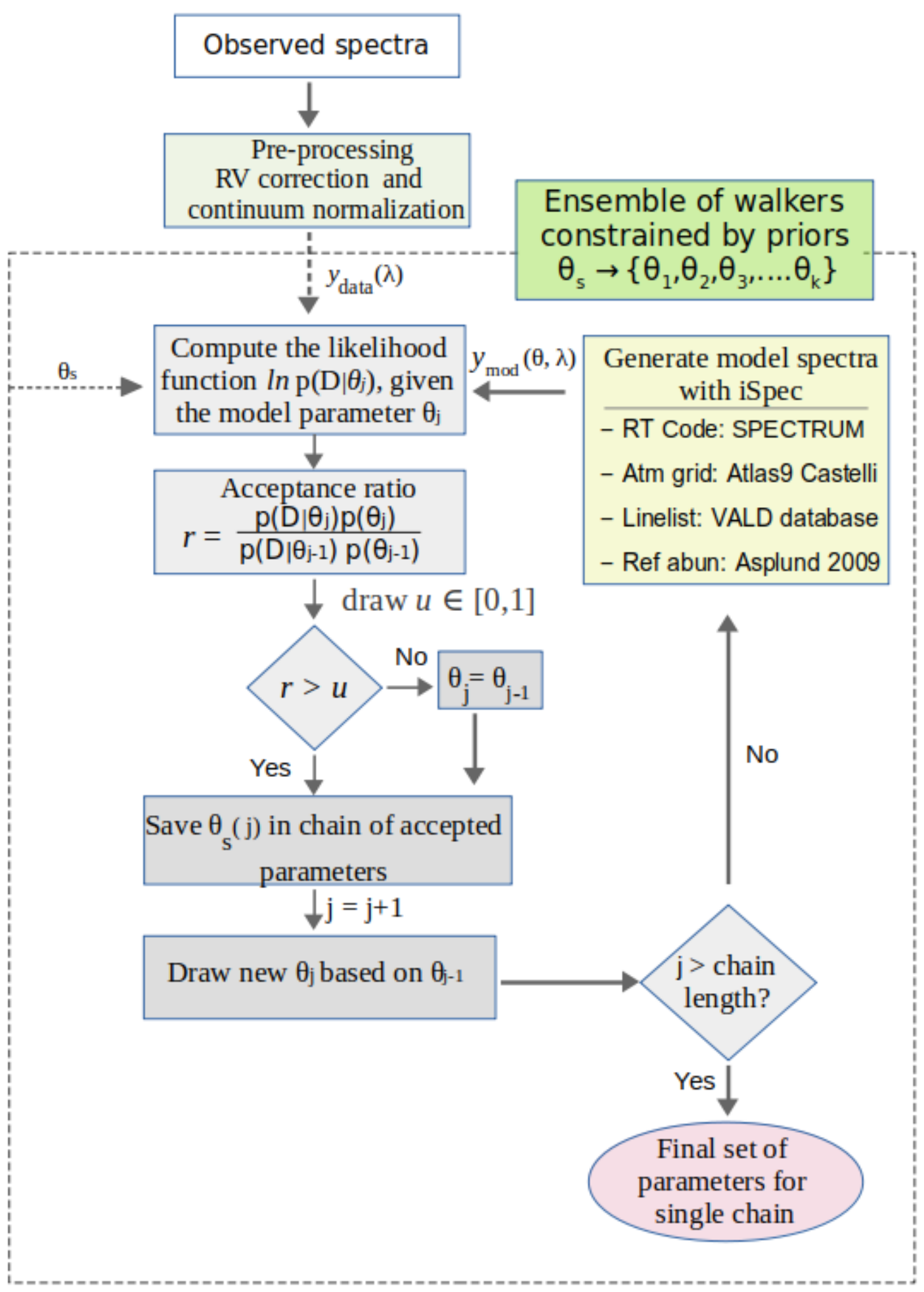}
\caption{Workflow diagram of iSpec along with {\emph{emcee}}. For a requested set of stellar parameters, iSpec generates the synthetic spectrum and compares it with the original spectrum. The most likely posterior distribution of the stellar parameters is obtained using Bayesian inference.}
\label{flowchart}
\end{figure}
We used {\emph{emcee}} implementation of MCMC\footnote{ https://emcee.readthedocs.io/en/stable/} in python. The flowchart of the our algorithm is  shown in Figure~\ref{flowchart}. First, we initialize the starting parameters $\theta_s$  of the model from our prior knowledge of the star, e.g. spectral type, luminosity class etc. Using $\theta_s$ as seed we generate an ensemble of $\{\theta_1, \theta_2,...\theta_k\}$ called walkers drawn from a physically realistic range of uniform priors, i.e. $\pm 200$~K for $T_\mathrm{eff}$ , $\pm 0.5$~dex for $\log\,g$,  $\pm 0.25$~dex for [Fe/H] and $\pm2$ to $\pm20$  for $v\cdot\sin i$ depending on the star. 

\begin{figure}[t!]
\includegraphics[width=3.5in]{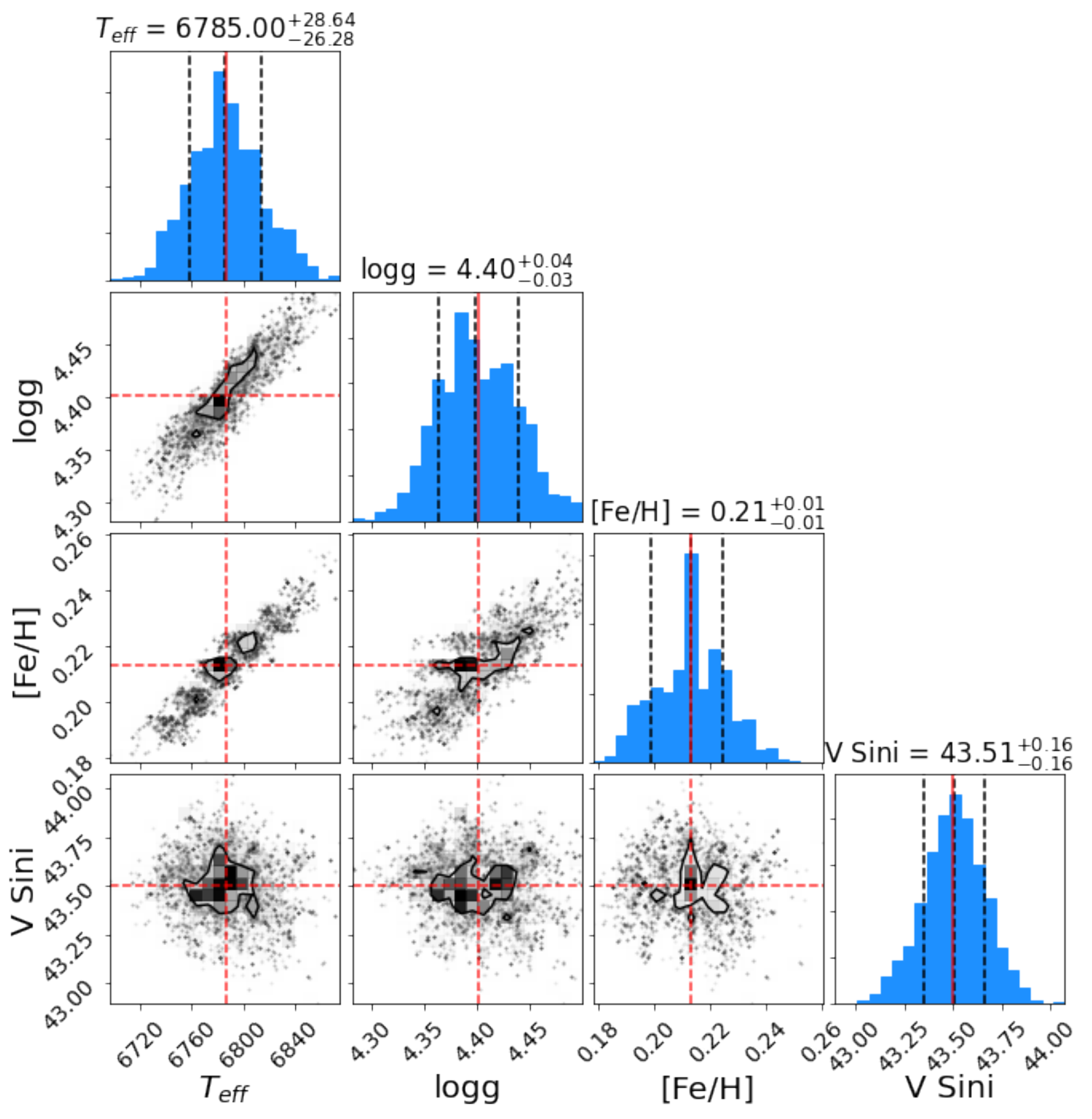}
\caption{Posterior distributions of stellar parameters for 
HR2562, obtained from MCMC analysis (40 chains, 300 steps, a burn-in limit at 140 steps). The diagonal panel shows 1-D projections of the probability density, while the off-diagonals show 2-D projections of the correlations between parameters. The mean of each parameter is shown by the red dashed lines, while the $1\sigma$ spread is indicated by black dashed lines.}
\label{MCMC}
\end{figure}

Each walker is a random realization of $\theta$ which relies on algorithm(e.g., Metropolis-Hastings) for sampling the parameter space. A function call to iSpec generates the model spectrum for the proposal parameter from the MCMC sampler.  We define a simple log likelihood function $lnP(D|\theta)$ to compares the observed spectrum $y_{\mathrm{data}}$  with the model spectrum $y_{\mathrm{mod}}$  as:

\begin{equation}
lnP(D|\theta)= - \frac{1}{2}\sum\Big(\frac{y_{\mathrm{data}}-y_{\mathrm{mod}}(D|\theta)}{y_{\mathrm{err}}}\Big)^2
\label{e2}
\end{equation}

Every walker numerically explores the parameter space by taking a ``step" to a new value $\theta_{j+1}$ that is drawn from a normal proposal distribution centered on $\theta_{j}$. The new proposal $\theta_{j+1}$ is accepted if it has a higher posterior value than the current sample, $\theta_{j}$. If the new proposal value has a lower posterior, then the choice to accept or reject a new proposal with a certain probability is made randomly.   

The walker, thus, guided by Markov's process, iteratively converges towards the target distribution by producing a chain of accepted parameters, as illustrated in Figure~\ref{flowchart}. We discard some of the early samples in each chain as they are likely to lie outside the target distribution. This is termed as ``burn-in". Finally, after the burn-in, we obtain a posterior distribution of our stellar parameters.
\begin{figure}[ht!]
\includegraphics[width=3.5in]{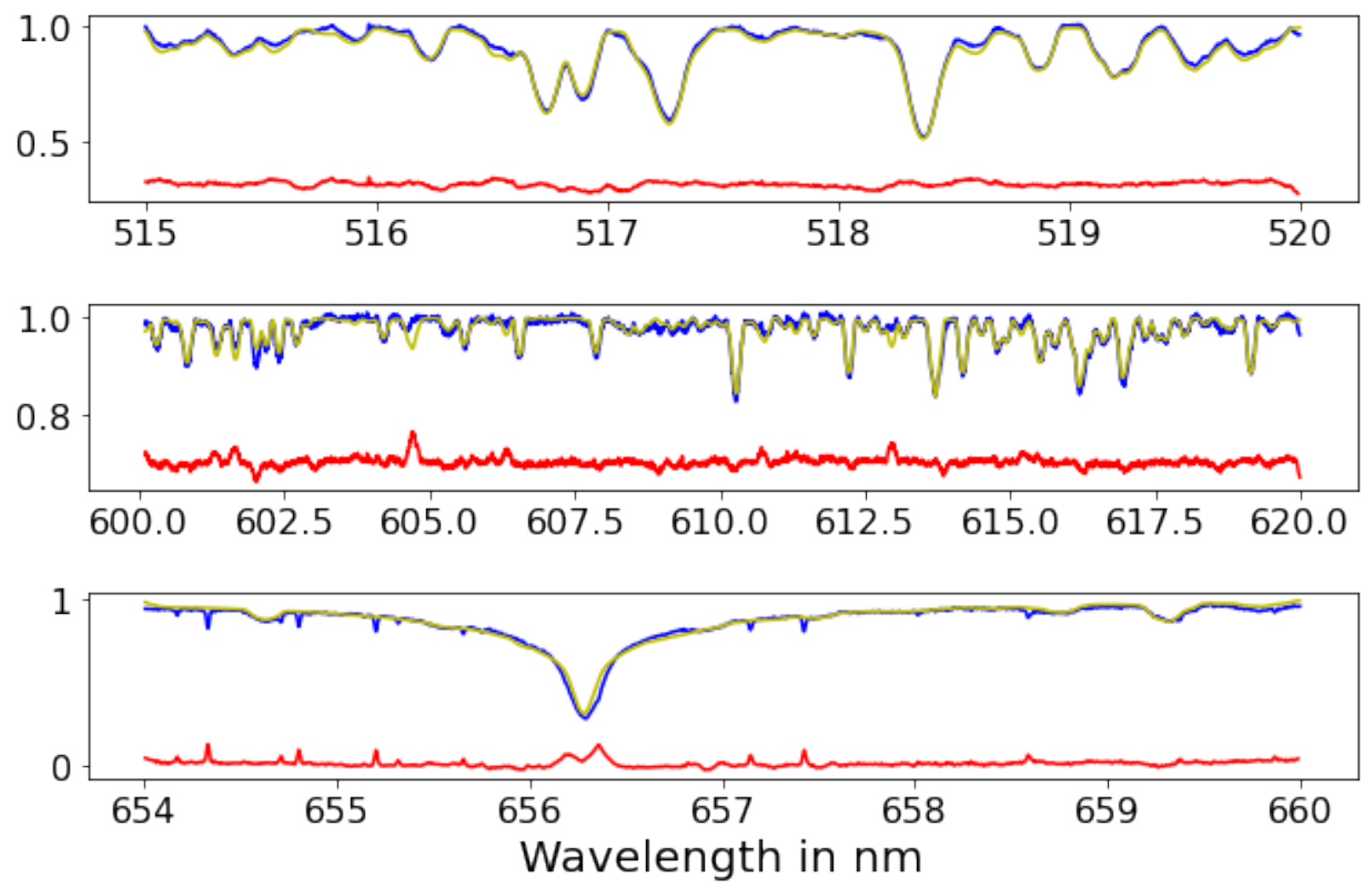}
\caption{Comparison between observed spectra (blue) and synthetic spectra (yellow) for  HR2562 in three distinct wavelength regions. The synthetic spectra were generated from stellar parameters obtained using Bayesian analysis. Note that offset is added to the residuals (red) for clarity.}
\label{spec-comp}
\end{figure}

\begin{figure}[b]
\includegraphics[width=3.5in]{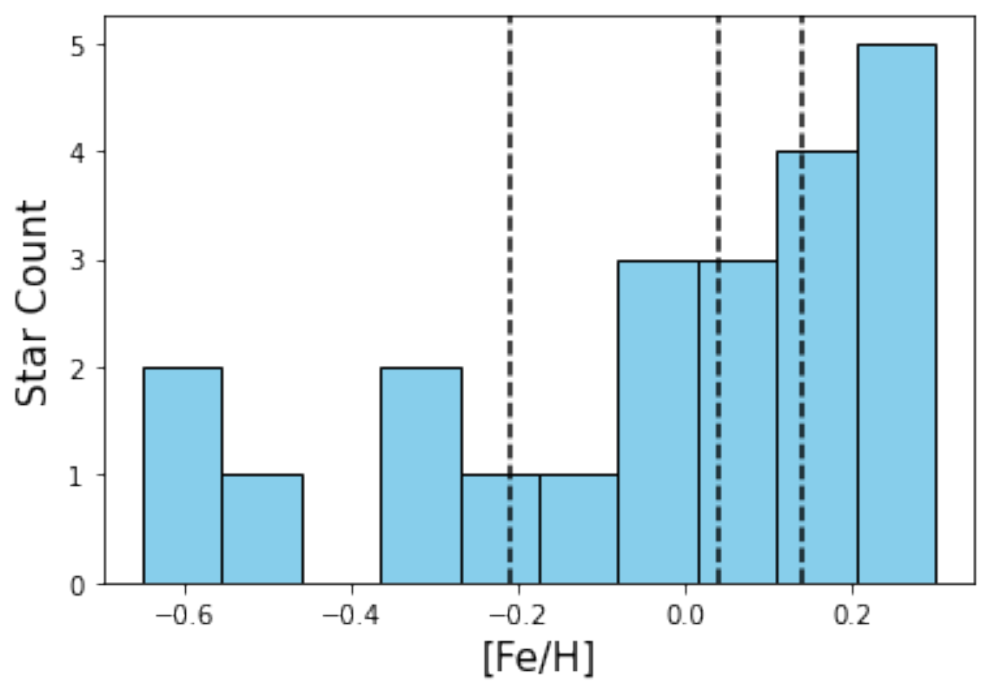}
\caption{The observed metallicity ([Fe/H]) distribution of a subset of stars (S.No. 1-22 in Table~\ref{star_params}.) known to host directly imaged planets. The dashed lines represent the median and the 1st and 3rd quartiles of the distribution.}
\label{metalicity}
\end{figure}

After some experimentation, we found that by using 300 steps following a burn-in limit of 140 steps for 40 test chains, we get a reasonable posterior distribution to determine the statistics of stellar parameters. For illustration, the final distribution of $T_{\mathrm{eff}}$, $\log\,g$, [Fe/H] and $v\cdot\sin i$ for HR2562 is shown in the Figure~\ref{MCMC}. Since our posterior distribution is multivariate, some of the model parameters are likely to correlate. The shape of the contour plots in Figure~\ref{MCMC} reflects the degree of correlation between different stellar parameters, e.g., expected correlation can be seen between $\log\,g$, and $T_{\mathrm{eff}}$ while for others; the scatter is uniform, implying no correlation. As a representative example, we show the synthetic spectra for HR 2562 generated using Bayesian inferred model parameters in Figure~\ref{spec-comp}, which matches reasonably well with the observed spectra. 

For the stars with veiling (S.No 13-18  in  Table~\ref{star_params}), the estimation of stellar parameters was done in parallel with determining of veiling. This was possible because the line shapes and relative absorption line depths are affected by the stellar parameters and are independent of the presence of veiling. We followed a similar procedure as described in \cite{2002A&A...391..595S,2003A&A...408..693S}, where veiling was modeled as free parameter $V(\lambda)$ in the log-likelihood function in \textbf{Eq.~\ref{e2}}. We used the modified log-likelihood function to obtain the stellar parameters by the same procedure as described above.

The final stellar parameters for our selected stars with mean values and $\pm 1 \sigma$ uncertainty are listed in Table~\ref{star_params}. The errors associated with the stellar parameters are the Bayesian error bars that are related to the sampling of the model spectra. The intrinsic uncertainty associated with the model generating the spectrum is not taken into account. Typical standard errors  associated with metallicity ($\pm 0.15$) are discussed in details by \cite{bla14} and \cite{jof19}.

\begin{figure}[t]
\includegraphics[width=3.6in]{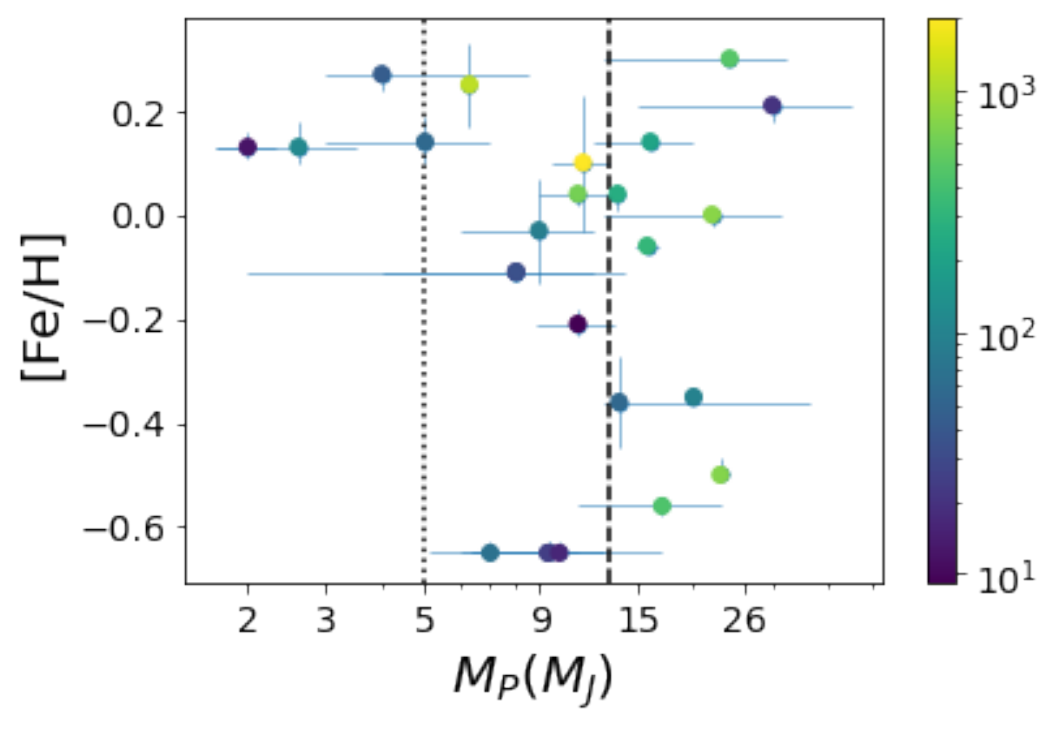}
\caption{The distribution of mass of directly imaged planets and the host-star metallicity. The dotted-line indicates  $5 M_J$ and dashed-line indicates $13M_J$ boundary. The color bar to the right represents the orbital distance in AU}  
\label{mass-metal}
\end{figure}

\section{Results} 
\begin{figure*}[t!]
\centering
\includegraphics[width=6.4in]{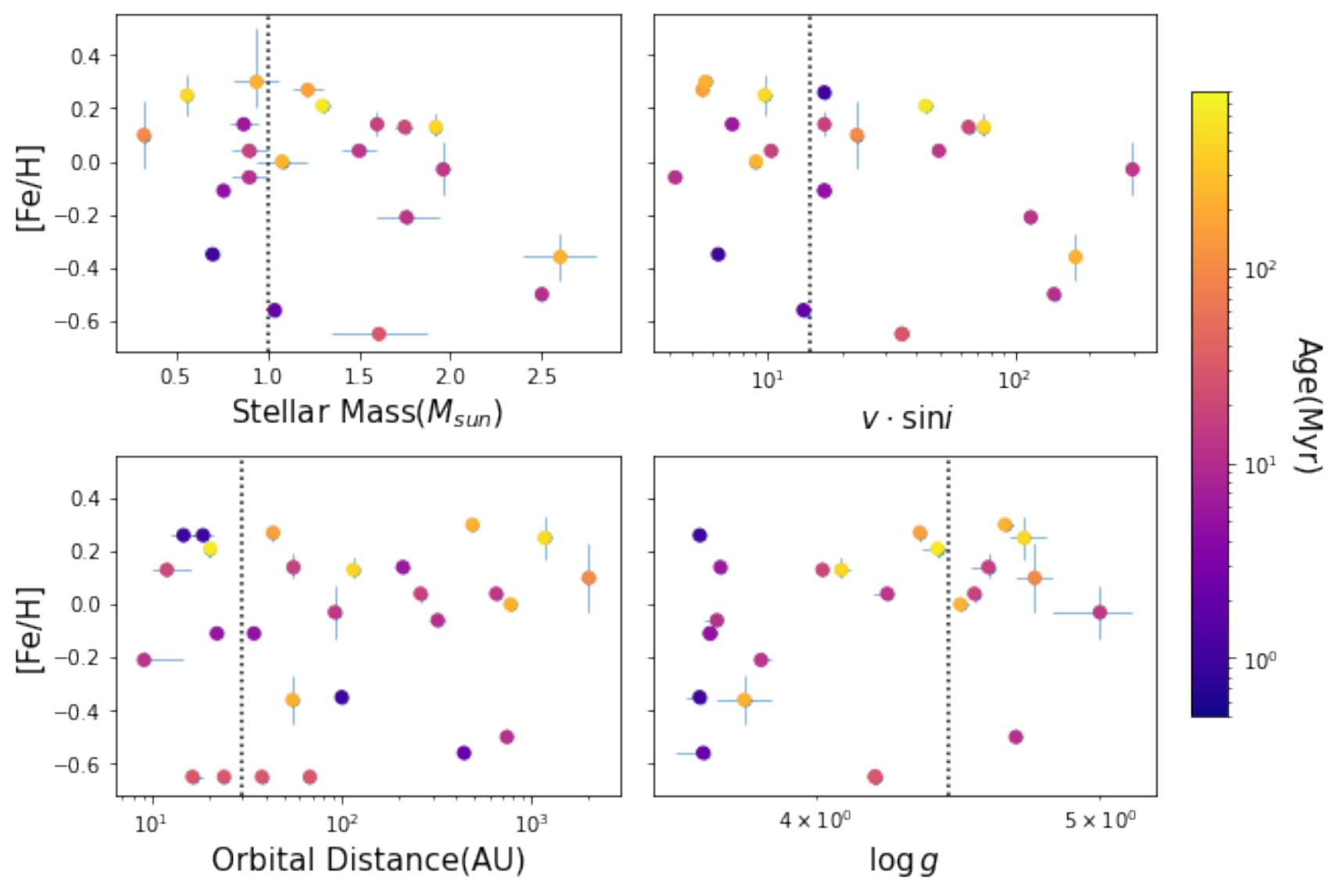}
\caption{Correlation of host-star metallicity with other stellar parameters. The dotted line in the bottom-left plot represents the orbital distance of the Neptune while in other plots it refers to corresponding parameters of the Sun.}
\label{metal-param}
\end{figure*}
\subsection{Metallicity of DIP host stars}
We have estimated the stellar parameters for a subset of stars harboring directly imaged planets listed in Table~\ref{star_params}. Figure~\ref{metalicity} shows the distribution of observed metallicity for 22 stars, 18 of which are analyzed in the present work, and the metallicity value for four stars is taken from previous studies. The metallicity of these targets do not show any trend or clustering but widely varies from +0.30~dex (HD 203030) to -0.65~dex (HR 8977) with median centered at 0.04~dex, which is closer to the solar value. The first and third quartiles are -0.21 and 0.14~dex, respectively, with 12 of them having metallicity higher than the solar value. The large scatter seen in [Fe/H] is not very surprising as it likely reflects the heterogeneity of the DIP host stars associated with different star-forming regions, parent clusters, or the moving groups.

\subsection{Metallicity and  planet mass}
To study the relationship between host-star metallicity and planet mass, we used the planetary mass data from the NASA Exoplanet Database from the composite planet list. We divided our DIP sample into three mass bins: $1M_{J}\!< \!M_p\! \leqslant \!5M_{J}$, $5M_{J}\! <\! M_p \!\leqslant 13M_{J}$ and $ M_p\!>\!13\! M_{J}$ as shown in Figure~\ref{mass-metal}. The average metallicity is $0.17\pm0.07$~dex for four stars in the 1st bin,  $-0.08\pm0.29$~dex for seven stars in the 2nd bin, and  $-0.11\pm 0.30$~dex for ten stars in the 3rd bin. The mean metallicity in each bin shows a declining trend with increasing planetary mass. We also note that regardless of their orbital distance, directly imaged planets with $M_{p}\leqslant 5M_{J}$ have mostly metal-rich hosts. 

\subsection{Metallicity and other stellar parameters}
Figure~\ref{metal-param} shows the distribution of metallicity as a function of orbital distance, stellar mass, $\log g $ and $ v\cdot\sin i $. For low-mass stars, $M_{\star}\le 1M_{\odot}$, we find that average metallicity is near-solar with standard deviation $0.21$~dex. Stars with $M_{\star}> 1M_{\odot}$ are found to be slightly metal-poor with average metallicity to be $-0.10$~dex and standard deviation $0.30$~dex.

We also find that the average metallicity of fast rotating stars  ($v\cdot\sin i> 15$~km/s) is $-0.1$~dex with a standard deviation of $0.29$~dex, while for slow rotators ($v\cdot\sin i< 15$~ km/s) it is solar, 0.02~dex with a standard deviation of 0.28~dex. The Spearman's rank correlation coefficient between the stellar metallicity and projected rotational velocity of the star $v\cdot\sin i$ is -0.42 with a p-value of 0.05, which suggests a weak negative correlation. Furthermore, there is no noticeable dependence of host star metallicity on orbital distance and $\log g$

\subsection{Comparison with literature}
To compare our results in Table.\ref{star_params}, we have included the stellar parameters of DIP host stars from the literature. For each stellar parameter, we computed the sample mean difference and the maximum deviation between our values and those reported in the literature. For effective temperature, we find the sample mean difference to be +103K and maximum deviation to be 380K for Lkca~15. We note that $T_{\textrm{eff}}$ for most hosts stars in literature is determined photometrically, which could account for the observed differences. For surface gravity, the sample mean difference is -0.06~dex, and the maximum difference is 0.58~dex for the HD~95086.  Likewise, for metallicity, the sample mean difference is found to be -0.035~dex and the maximum difference, seen again for HD~95086, is 0.39~dex.  For rotation velocity, we find a good match between the literature and our values for slowly rotating DIP hosts ($v\cdot\sin i < 20$), whereas the maximum difference is found to be about 16~km/s for the fast rotating star Fomalhaut. By and large, our values for [Fe/H] and $\log g$ determined uniformly using the spectroscopic method are within the error margin of those quoted in the literature. However, for such a heterogeneous sample, the observed differences in stellar parameters obtained by different analysis methods, atmospheric models, radiative transfer codes and line lists, etc., are not entirely unexpected \citep{jof14, jof19, bla19}.
\begin{figure}
\gridline{\fig{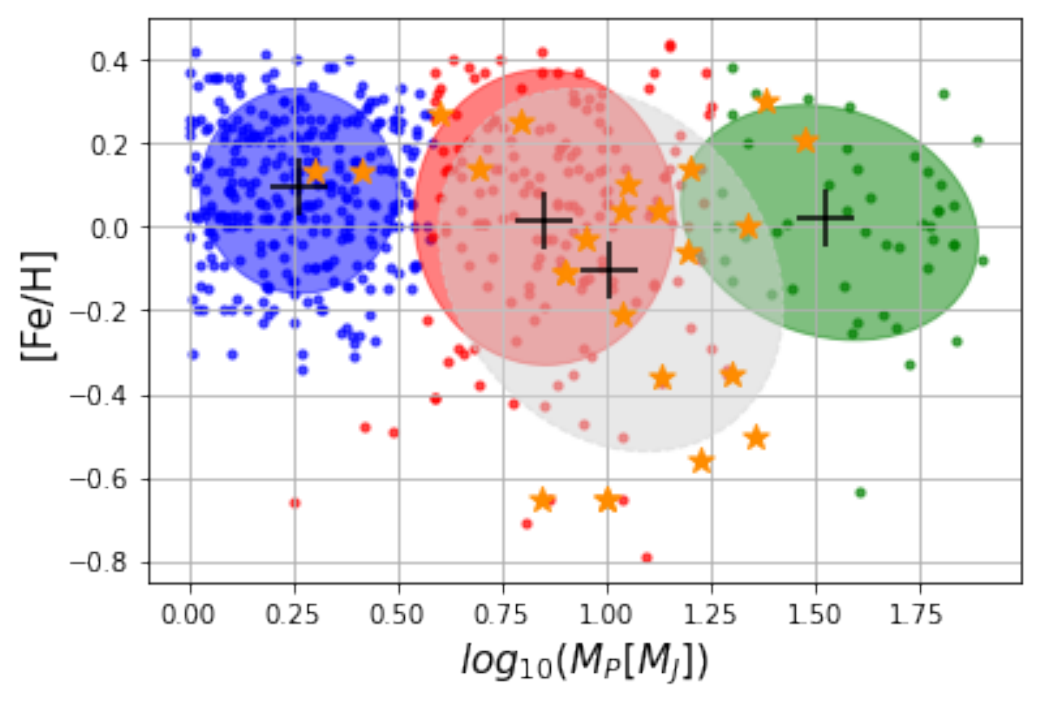}{0.5\textwidth}{}}
 \gridline{\fig{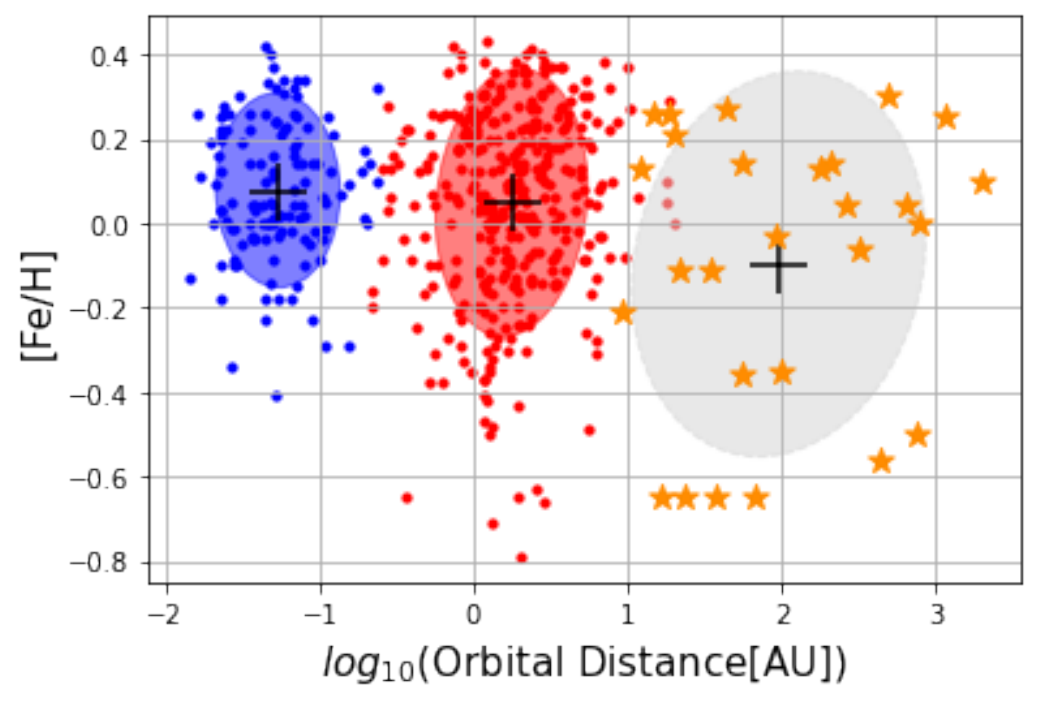}{0.5\textwidth}{}
          }
\caption{Top panel: A Gaussian mixture analysis of the combined sample of giant planets and brown-dwarf in \emph{metallicity-planetary mass} plane. Three separate clusters correspond to Jupiter-type  (blue), super-Jupiters (red), and brown-dwarfs (green). Bottom panel: Two populations resulting from the Gaussian mixture analysis in \emph{metallicity-orbital distance} plane (red and blue). 
The DIP population (orange $\star$ with a gray ellipse) analyzed in this work is interspersed between super-Jupiters and brown-dwarfs, as shown in the top panel in the metallicity-planetary mass plane, whereas it occupies a separate region in the metallicity- orbital distance plane. The centroid of each cluster is indicated by '+' symbols. 
\label{cluster-mass-od}}
\end{figure}
\section{Discussion}
In the standard paradigm for the formation of a Jupiter-like planet via core nucleated accretion \citep[e.g.][]{2018haex.bookE.140D}, a rocky protoplanetary core forms first, which then accretes gas and dust from the surrounding disk to become a gas giant \citep{Boss1836,1986Icar...67..391B,Pollack96,2001ApJ...553..999I}. The critical (or minimum) core mass required to form a gas giant depends on various factors (e.g., location on the protoplanetary disk, accretion rate of solids, etc.) and generally decreases with increasing disk radius: minimum core mass drops from $\sim$~8.5~$M_\oplus$ at 5 AU to $\sim$~3.5~$M_\oplus$ at 100 AU \citep{2014ApJ...786...21P, 2015ApJ...800...82P}. If the protoplanetary disk is rich in solids, i.e., higher metallicity, then the rocky core can grow faster and reach the critical mass for gas accretion well before the disk is depleted of gas. Therefore, it is easier to form Jupiter-like gas giants in disks around higher metallicity stars \cite[e.g.,][]{Ida04, Kornet05, Wyatt07, Boss10, 2012A&A...541A..97M}. Indeed, observations have shown that the frequency of Jupiter-like planets is higher around higher metallicity stars \citep[e.g.][]{1997MNRAS.285..403G, san01, fis05, 2007ARA&A..45..397U}. While not as strong as that seen for gas giants, smaller planets also show a weaker tendency to occur more frequently around relatively higher metallicity stars, even though their host stars appear to have a larger spread in the metallicity  \citep[e.g.,][]{wan15, buc14, Mul16}. It has now been adequately established that the host star metallicity ([Fe/H]), on average, increases with increasing planet mass or radius \citep[e.g.][]{buc14, pet18, nar18, mul18}.
 Thus the observed strong dependence of the planet mass/radius on the host star metallicity supports the core accretion model for planet formation.
\begin{figure}
\includegraphics[width=3.3in]{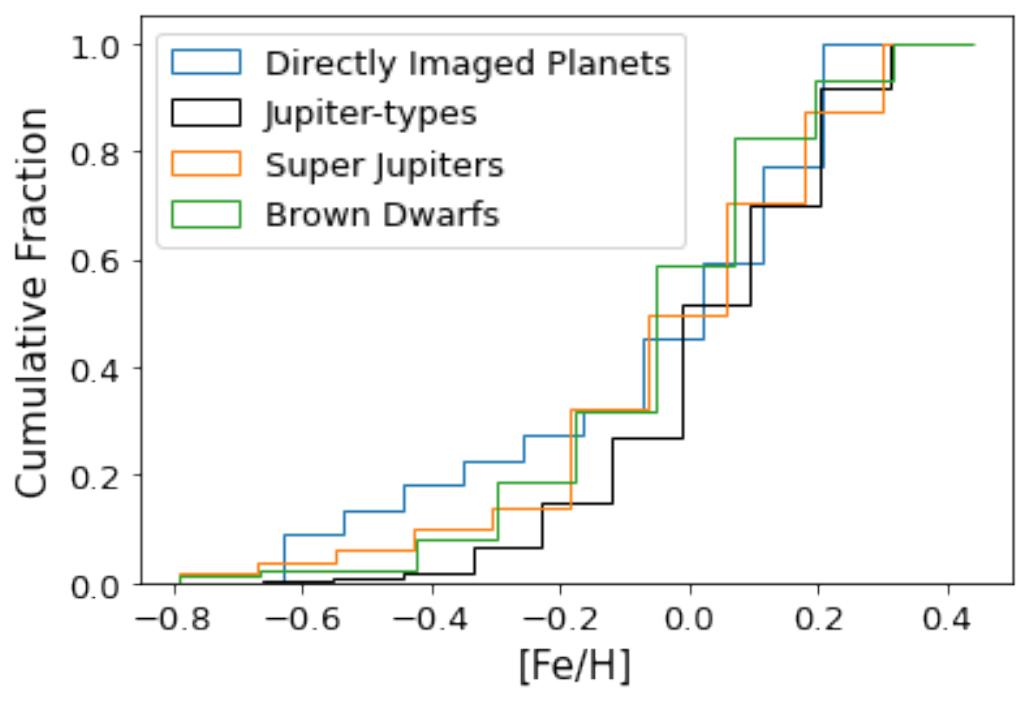}
\caption{Cumulative metallicity distribution of DIP host stars from the present studied (blue). The green curve represents cumulative metallicity  distribution of brown-dwarf companions while the black and orange curve represents the Jupiter-type and super Jupiter's respectively.}
\label{mass-metal-lit}
\end{figure}
However, the observed correlation of increasing host star metallicity with increasing planet mass turns over at about 4-5 $M_J$. For planet masses higher than this (super-Jupiters), the correlation reverses, and the average host star metallicity decreases as the mass of the planet increases \citep{san17,nar18}. This suggests that stars hosting super-Jupiters are not necessarily metal-rich, unlike stars hosting Jupiters. This trend appears to continue for more massive companions: the average metallicity of stars with a brown-dwarf secondary is also close to solar to sub-solar, and not super-solar like stars hosting Jupiters. \citep{ma14,nar18,2018ApJ...853...37S}. 

Our sample of directly imaged planets occupy a mass range similar to that of super-Jupiters and brown-dwarfs. The fact that the average host star metallicities of brown-dwarfs and super-Jupiters are similar and that they differ from that of Jupiter-hosts perhaps indicates a similar formation scenario for them that is different from that of Jupiters. It has been suggested that massive planets and low-mass brown-dwarfs can form via gravitational fragmentation of the disk rather than core accretion \citep[e.g.][]{Boss1836, 2002Sci...298.1756M}.  This gravitational instability model of planet formation predicts no dependence between planet mass and host star metallicity  \citep[e.g.,][]{Boss02, 2006ApJ...636L.149C, 2007ApJ...662.1282M, Boss10} unlike the core accretion model that predicts such a dependence. 

We further compare the directly imaged planets with the large population of giant planets and brown-dwarfs around main-sequence stars discovered by techniques other than the direct imaging. To this end, we found 637 stars hosting 746 giant planets and massive objects with mass range $1-55 M_\mathrm{J}$ listed in NASA's exoplanet archive. We also searched the above sample in the SWEET-CATALOG \citep{san13, sau18}, which provides the metallicity information for 459 stellar hosts having 494 companions.  Additionally, a catalog of 58 brown-dwarfs and their stellar companions was chosen from \cite{ma14}. A joint sample of 552 objects was formed by combining the giant planets and brown dwarfs. This combined sample has a mass range from $1- 80 M_\mathrm{J}$ and orbital distance spanning 0.02-20 AU. Since objects in the combined sample come from  RV, transits, TTV, astrometry, and microlensing observations, we have used the minimum mass ($M\cdot\sin i $) wherever the true mass was not available. 

We then ran a clustering analysis on the 2D-data set of combined samples of giant planets and brown-dwarfs with host star metallicity as one parameter and orbital-distance and companion-mass as other.  For clustering analysis, we considered a Gaussian mixture model and implemented using a Python library \emph{scikit-learn} package \citep{scikit-learn}. The Gaussian mixture model optimally segregated the combined sample into three clusters in \emph{metallicity - planet mass} plane, as shown in the top panel of  Figure~\ref{cluster-mass-od}  and into two clusters in \emph{metallicity - orbital distance} plane as shown in the bottom panel of the Figure~\ref{cluster-mass-od}.

The clustering analysis in Figure~\ref{cluster-mass-od} at the top clearly divides the combined sample into three mass and metallicity bins. The mass boundaries roughly located at $\approx 4M_\mathrm{J}$ and $\approx 14M_\mathrm{J}$ are consistent with multiple population of giant planets (i.e., Jupiters and super-Jupiters) and brown-dwarfs, pointing to their different physical origin. Further on, the declining centroid metallicity of each group in Figure~\ref{cluster-mass-od} at the top, i.e., $0.089\pm0.02$, $0.023\pm0.002$ to  $0.013\pm0.009$~dex,  is also consistent with previous results. The DIP population studied in this work is also shown for comparison in Figure~\ref{cluster-mass-od}. The DIP population falls between the super-Jupiters and brown-dwarfs population, both in mass and metallicity.

The analysis of orbital-distance and stellar-metallicity shows that the combined population of close-in objects separates into two distinct groups, as shown in the bottom panel of the Figure~\ref{cluster-mass-od}. Again, the DIP sample analyzed in this work is added to the plot for comparison. In \emph{metallicity -- orbital distance} plane, three populations again clearly separate out. On comparing the centroid values of the metallicity, which are 0.076, 0.042, and -0.097~dex, (standard deviation in each case  $\le10^{-6}$), we find a decreasing metallicity trend with increasing orbital distance. A similar metallicity dependence with orbital distance is also reported for the Jupiter analogs \citep{Mul16,buc18,mul18}. 

In Figure~\ref{mass-metal-lit}, we compare the cumulative metallicity distribution of DIP host stars with stellar companions of brown-dwarfs  \cite{ma14}, and giant planets --both Jupiter-type and super-Jupiters.  We note that the cumulative distribution of DIP host stars at the lower metallicity region clearly differs from the stellar hosts of Jupiter-type planets, whereas the distribution for super-Jupiters and brown-dwarf hosts is falling in between the two. However, there is no marked difference in the higher metallicity side beyond [Fe/H]$>0$.

Although the specific factors that influence planet formation are still not fully understood, metallicity seems to be one of the major contributing factors which determine the type of planets likely to be formed around a star. Using synthetic planet population models \cite{2012A&A...541A..97M} showed that a high-metallicity environment determines whether or not a giant planet in the mass range $1-4M_\mathrm{J}$ can form. But metallicity alone is not the only parameter in determining the final mass of the planet except for the very massive planets ($\geq 10M_\mathrm{J}$), as the critical core must form very fast before the dissipation of the gas in the disk by accretion onto the star \citep{1985prpl.conf.1100H,2007ApJ...662.1282M}.  The prediction of  \cite{2012A&A...541A..97M} that the very massive planets ($\geq$ $10M_{\mathrm{J}}$) can form only at very high metallicity conditions is contrary to our findings. Our results are indicative of the possibility of two planet formation pathways: one in which the giant planets up to $4-5M_{J}$  might are formed by core accretion process, and the other where the massive super-Jupiters and brown-dwarfs are formed via gravitational fragmentation of the protoplanetary disk.

Our results for wide-orbit (10s-1000s~AU) planets are also consistent with the mass-metallicity trend observed for super-Jupiters and brown-dwarfs in close-in ($\la$~1 AU) orbits around main-sequence stars. The formation mechanism of planets in wider orbits is still unclear. However, the mixed metallicity of our DIP host star sample and its close resemblance with commutative metallicity distribution of brown-dwarf hosts make it likely that massive and young planets in wider-orbits too formed via gravitational instability. However,  a larger sample is required to further validate such conclusions.

 \section{Summary and conclusions} 
We have used high-resolution spectra to measure the atmospheric parameters of young stars that are confirmed host stars of planets detected by direct imaging technique. Our sample consists of 22 such stars selected from NASA's Exoplanet Archive. For 18 of these targets, the stellar parameters and metallicity are determined in a uniform and consistent way. The summary of our results is as follows:

\begin{enumerate}

\item We used the Bayesian analysis to estimated the atmospheric parameters and metallicity for 18 DIP host stars. The MCMC technique was used to obtain the posterior distribution of stellar using model spectra generated using the iSpec. The computed metallicity [Fe/H] of these stars spans a wide range from between $+0.3$ and $-0.65$~dex. 

\item We investigated the trend between the average host star metallicity and mass of the planet, which shows that directly imaged planets with $M_{P}\leqslant 5M_{J}$ tend to have metal-rich hosts. This is in line with the predictions of planet formation via core accretion. However, as the planet mass increases,  the average metallicity of the host stars shows a declining trend, suggesting that these planets are likely formed by gravitational instability. These findings seem consistent with the results reported by \citet{san17} and \citet{nar18}. Since  the metallicity of a star doesn't change during evolution, we do not expect these trends to change significantly for the currently undetected population of cool and massive giant planets in the outstretched regions of the main sequence stars.
Moreover, main sequence host stars, in general,  show a trend of decreasing metallicity with increasing orbital distance of the planet (e.g., \cite{Mul16}, \cite{buc18}, \cite{mul18}, \cite{nar18}). 

\item From clustering analysis, as discussed above in section 6, we find that the DIP host stars separate as a different class of celestial objects in stellar metallicity--orbital distance plane. Furthermore, we can see a decreasing trend in the centroids of the host-star metallicity as the star-planet separation increases.

\item In the planetary mass- stellar metallicity plane, it is found that the Jupiter-like planets are more likely to form around a metal-rich star. It also shows a decreasing trend in average stellar metallicity as the planetary mass increases. The DIP population clusters lie in between the super -Jupiters and brown dwarf populations.
\end{enumerate}

It is also important to recognize that the composition of circumstellar material from which the planets are formed needn't necessarily be the same as the composition of the parent star. The degree of similarity or difference would depend on how and where planets are formed, what stage of evolution they are in, the disk mass and planet multiplicity. A clear picture is expected to emerge from the ongoing high contrast imaging surveys and future experiments aimed at searching planets in wider orbits.

\section*{ACKNOWLEDGMENTS}
This work has made use of (a) ESO archival data that was observed under programs 192.C-0224, 098.C-0739, 266.D-5655, 094.A-9012, 084.C-1039, 074.C-0037, and 65.I-0404; (b) the Keck Observatory Archive (KOA), which is operated by the W. M. Keck Observatory and the NASA Exoplanet Science Institute (NExScI), under contract with the National Aeronautics and Space Administration; (c)  the NASA Exoplanet Database, which is run by the California Institute of Technology under an Exoplanet Exploration Program contract with the National Aeronautics and Space Administration and (d) the European Space Agency (ESA) space mission Gaia, the data from which were processed by the Gaia Data Processing and Analysis Consortium (DPAC). Funding for the DPAC has been provided by national institutions, in particular the institutions participating in the Gaia Multilateral Agreement.  

Some initial test observations for this work were taken with 2m Himalayan Chandra Telescope at Hanle. The facilities at IAO and CREST are operated by the Indian Institute of Astrophysics, Bangalore. Furthermore, this work has made use of the NASA Exoplanet Database, which is run by the California Institute of Technology under an Exoplanet Exploration Program contract with the National Aeronautics and Space Administration. Swastik C.  would also like to thank Aritra Chakrabarty at the Indian Institute of Astrophysics for the insightful discussion on the Bayesian analysis.
\\

\software{astropy \citep{astropy:2013},  
          iSpec \citep{bla14}, 
          emcee \citep{2013PASP..125..306F},
          scikit-learn \citep{scikit-learn}, 
          IRAF \citep{1986SPIE..627..733T, 1993ASPC...52..173T}
                    }.

\bibliography{biblio}{}

\begin{thebibliography}{}
\expandafter\ifx\csname natexlab\endcsname\relax\def\natexlab#1{#1}\fi
\providecommand{\url}[1]{\href{#1}{#1}}
\providecommand{\dodoi}[1]{doi:~\href{http://doi.org/#1}{\nolinkurl{#1}}}
\providecommand{\doeprint}[1]{\href{http://ascl.net/#1}{\nolinkurl{http://ascl.net/#1}}}
\providecommand{\doarXiv}[1]{\href{https://arxiv.org/abs/#1}{\nolinkurl{https://arxiv.org/abs/#1}}}

\bibitem[{{Adibekyan}(2019)}]{Adibekyan19}
{Adibekyan}, V. 2019, Geosciences, 9, 105, \dodoi{10.3390/geosciences9030105}

\bibitem[{{Akeson} {et~al.}(2013){Akeson}, {Chen}, {Ciardi}, {Crane}, {Good},
  {Harbut}, {Jackson}, {Kane}, {Laity}, {Leifer}, {Lynn}, {McElroy}, {Papin},
  {Plavchan}, {Ram{\'\i}rez}, {Rey}, {von Braun}, {Wittman}, {Abajian}, {Ali},
  {Beichman}, {Beekley}, {Berriman}, {Berukoff}, {Bryden}, {Chan}, {Groom},
  {Lau}, {Payne}, {Regelson}, {Saucedo}, {Schmitz}, {Stauffer}, {Wyatt}, \&
  {Zhang}}]{2013PASP..125..989A}
{Akeson}, R.~L., {Chen}, X., {Ciardi}, D., {et~al.} 2013, \pasp, 125, 989,
  \dodoi{10.1086/672273}

\bibitem[{{Artigau} {et~al.}(2015){Artigau}, {Gagn{\'e}}, {Faherty}, {Malo},
  {Naud}, {Doyon}, {Lafreni{\`e}re}, \& {Beletsky}}]{2015ApJ...806..254A}
{Artigau}, {\'E}., {Gagn{\'e}}, J., {Faherty}, J., {et~al.} 2015, \apj, 806,
  254, \dodoi{10.1088/0004-637X/806/2/254}

\bibitem[{{Asplund} {et~al.}(2009){Asplund}, {Grevesse}, {Sauval}, \&
  {Scott}}]{2009ARA&A..47..481A}
{Asplund}, M., {Grevesse}, N., {Sauval}, A.~J., \& {Scott}, P. 2009, \araa, 47,
  481, \dodoi{10.1146/annurev.astro.46.060407.145222}

\bibitem[{{Astropy Collaboration} {et~al.}(2013){Astropy Collaboration},
  {Robitaille}, {Tollerud}, {Greenfield}, {Droettboom}, {Bray}, {Aldcroft},
  {Davis}, {Ginsburg}, {Price-Whelan}, {Kerzendorf}, {Conley}, {Crighton},
  {Barbary}, {Muna}, {Ferguson}, {Grollier}, {Parikh}, {Nair}, {Unther},
  {Deil}, {Woillez}, {Conseil}, {Kramer}, {Turner}, {Singer}, {Fox}, {Weaver},
  {Zabalza}, {Edwards}, {Azalee Bostroem}, {Burke}, {Casey}, {Crawford},
  {Dencheva}, {Ely}, {Jenness}, {Labrie}, {Lim}, {Pierfederici}, {Pontzen},
  {Ptak}, {Refsdal}, {Servillat}, \& {Streicher}}]{astropy:2013}
{Astropy Collaboration}, {Robitaille}, T.~P., {Tollerud}, E.~J., {et~al.} 2013,
  \aap, 558, A33, \dodoi{10.1051/0004-6361/201322068}

\bibitem[{{Bailey} {et~al.}(2014){Bailey}, {Meshkat}, {Reiter}, {Morzinski},
  {Males}, {Su}, {Hinz}, {Kenworthy}, {Stark}, {Mamajek}, {Briguglio}, {Close},
  {Follette}, {Puglisi}, {Rodigas}, {Weinberger}, \&
  {Xompero}}]{2014ApJ...780L...4B}
{Bailey}, V., {Meshkat}, T., {Reiter}, M., {et~al.} 2014, \apjl, 780, L4,
  \dodoi{10.1088/2041-8205/780/1/L4}

\bibitem[{{Baraffe} {et~al.}(2010){Baraffe}, {Chabrier}, \& {Barman}}]{bar10}
{Baraffe}, I., {Chabrier}, G., \& {Barman}, T. 2010, Reports on Progress in
  Physics, 73, 016901, \dodoi{10.1088/0034-4885/73/1/016901}

\bibitem[{{Baraffe} {et~al.}(2003){Baraffe}, {Chabrier}, {Barman}, {Allard}, \&
  {Hauschildt}}]{bar03}
{Baraffe}, I., {Chabrier}, G., {Barman}, T.~S., {Allard}, F., \& {Hauschildt},
  P.~H. 2003, \aap, 402, 701, \dodoi{10.1051/0004-6361:20030252}

\bibitem[{{Baron} {et~al.}(2019){Baron}, {Lafreni{\`e}re}, {Artigau},
  {Gagn{\'e}}, {Rameau}, {Delorme}, \& {Naud}}]{Baron19}
{Baron}, F., {Lafreni{\`e}re}, D., {Artigau}, {\'E}., {et~al.} 2019, \aj, 158,
  187, \dodoi{10.3847/1538-3881/ab4130}

\bibitem[{{Batalha}(2014)}]{bat14}
{Batalha}, N.~M. 2014, Proceedings of the National Academy of Science, 111,
  12647, \dodoi{10.1073/pnas.1304196111}

\bibitem[{{B{\'e}jar} {et~al.}(2008){B{\'e}jar}, {Zapatero Osorio},
  {P{\'e}rez-Garrido}, {{\'A}lvarez}, {Mart{\'\i}n}, {Rebolo},
  {Vill{\'o}-P{\'e}rez}, \& {D{\'\i}az-S{\'a}nchez}}]{2008ApJ...673L.185B}
{B{\'e}jar}, V.~J.~S., {Zapatero Osorio}, M.~R., {P{\'e}rez-Garrido}, A.,
  {et~al.} 2008, \apjl, 673, L185, \dodoi{10.1086/527557}

\bibitem[{{Blanco-Cuaresma}(2019)}]{bla19}
{Blanco-Cuaresma}, S. 2019, \mnras, 486, 2075, \dodoi{10.1093/mnras/stz549}

\bibitem[{{Blanco-Cuaresma} {et~al.}(2014{\natexlab{a}}){Blanco-Cuaresma},
  {Soubiran}, {Heiter}, \& {Jofr{\'e}}}]{bla14}
{Blanco-Cuaresma}, S., {Soubiran}, C., {Heiter}, U., \& {Jofr{\'e}}, P.
  2014{\natexlab{a}}, \aap, 569, A111, \dodoi{10.1051/0004-6361/201423945}

\bibitem[{{Blanco-Cuaresma} {et~al.}(2014{\natexlab{b}}){Blanco-Cuaresma},
  {Soubiran}, {Jofr{\'e}}, \& {Heiter}}]{Blac14}
{Blanco-Cuaresma}, S., {Soubiran}, C., {Jofr{\'e}}, P., \& {Heiter}, U.
  2014{\natexlab{b}}, \aap, 566, A98, \dodoi{10.1051/0004-6361/201323153}

\bibitem[{{Bochanski} {et~al.}(2018){Bochanski}, {Faherty}, {Gagn{\'e}},
  {Nelson}, {Coker}, {Smithka}, {Desir}, \& {Vasquez}}]{2018AJ....155..149B}
{Bochanski}, J.~J., {Faherty}, J.~K., {Gagn{\'e}}, J., {et~al.} 2018, \aj, 155,
  149, \dodoi{10.3847/1538-3881/aaaebe}

\bibitem[{{Bodenheimer} \& {Pollack}(1986)}]{1986Icar...67..391B}
{Bodenheimer}, P., \& {Pollack}, J.~B. 1986, \icarus, 67, 391,
  \dodoi{10.1016/0019-1035(86)90122-3}

\bibitem[{{Bohn} {et~al.}(2020){Bohn}, {Kenworthy}, {Ginski}, {Manara},
  {Pecaut}, {de Boer}, {Keller}, {Mamajek}, {Meshkat}, {Reggiani}, {Todorov},
  \& {Snik}}]{2020MNRAS.492..431B}
{Bohn}, A.~J., {Kenworthy}, M.~A., {Ginski}, C., {et~al.} 2020, \mnras, 492,
  431, \dodoi{10.1093/mnras/stz3462}

\bibitem[{{Bonnefoy} {et~al.}(2014){Bonnefoy}, {Currie}, {Marleau},
  {Schlieder}, {Wisniewski}, {Carson}, {Covey}, {Henning}, {Biller}, {Hinz},
  {Klahr}, {Marsh Boyer}, {Zimmerman}, {Janson}, {McElwain}, {Mordasini},
  {Skemer}, {Bailey}, {Defr{\`e}re}, {Thalmann}, {Skrutskie}, {Allard},
  {Homeier}, {Tamura}, {Feldt}, {Cumming}, {Grady}, {Brandner}, {Helling},
  {Witte}, {Hauschildt}, {Kandori}, {Kuzuhara}, {Fukagawa}, {Kwon}, {Kudo},
  {Hashimoto}, {Kusakabe}, {Abe}, {Brandt}, {Egner}, {Guyon}, {Hayano},
  {Hayashi}, {Hayashi}, {Hodapp}, {Ishii}, {Iye}, {Knapp}, {Matsuo}, {Mede},
  {Miyama}, {Morino}, {Moro-Martin}, {Nishimura}, {Pyo}, {Serabyn}, {Suenaga},
  {Suto}, {Suzuki}, {Takahashi}, {Takami}, {Takato}, {Terada}, {Tomono},
  {Turner}, {Watanabe}, {Yamada}, {Takami}, \& {Usuda}}]{BC}
{Bonnefoy}, M., {Currie}, T., {Marleau}, G.~D., {et~al.} 2014, \aap, 562, A111,
  \dodoi{10.1051/0004-6361/201322119}

\bibitem[{{Boro Saikia} {et~al.}(2015){Boro Saikia}, {Jeffers}, {Petit},
  {Marsden}, {Morin}, \& {Folsom}}]{2015A...573A..17B}
{Boro Saikia}, S., {Jeffers}, S.~V., {Petit}, P., {et~al.} 2015, \aap, 573,
  A17, \dodoi{10.1051/0004-6361/201424096}

\bibitem[{{Borucki} {et~al.}(2011){Borucki}, {Koch}, {Basri}, {Batalha},
  {Boss}, {Brown}, {Caldwell}, {Christensen-Dalsgaard}, {Cochran}, {DeVore},
  {Dunham}, {Dupree}, {Gautier}, {Geary}, {Gilliland}, {Gould}, {Howell},
  {Jenkins}, {Kjeldsen}, {Latham}, {Lissauer}, {Marcy}, {Monet}, {Sasselov},
  {Tarter}, {Charbonneau}, {Doyle}, {Ford}, {Fortney}, {Holman}, {Seager},
  {Steffen}, {Welsh}, {Allen}, {Bryson}, {Buchhave}, {Chandrasekaran},
  {Christiansen}, {Ciardi}, {Clarke}, {Dotson}, {Endl}, {Fischer}, {Fressin},
  {Haas}, {Horch}, {Howard}, {Isaacson}, {Kolodziejczak}, {Li}, {MacQueen},
  {Meibom}, {Prsa}, {Quintana}, {Rowe}, {Sherry}, {Tenenbaum}, {Torres},
  {Twicken}, {Van Cleve}, {Walkowicz}, \& {Wu}}]{bor11}
{Borucki}, W.~J., {Koch}, D.~G., {Basri}, G., {et~al.} 2011, \apj, 728, 117,
  \dodoi{10.1088/0004-637X/728/2/117}

\bibitem[{Boss(1997)}]{Boss1836}
Boss, A.~P. 1997, Science, 276, 1836, \dodoi{10.1126/science.276.5320.1836}

\bibitem[{{Boss}(2002)}]{Boss02}
{Boss}, A.~P. 2002, \apjl, 567, L149, \dodoi{10.1086/340108}

\bibitem[{{Boss}(2010)}]{Boss10}
{Boss}, A.~P. 2010, in IAU Symposium, Vol. 265, Chemical Abundances in the
  Universe: Connecting First Stars to Planets, ed. K.~{Cunha}, M.~{Spite}, \&
  B.~{Barbuy}, 391--398, \dodoi{10.1017/S1743921310001067}

\bibitem[{{Bowler}(2016)}]{bow16}
{Bowler}, B.~P. 2016, \pasp, 128, 102001,
  \dodoi{10.1088/1538-3873/128/968/102001}

\bibitem[{{Bowler} {et~al.}(2013){Bowler}, {Liu}, {Shkolnik}, \&
  {Dupuy}}]{2013ApJ...774...55B}
{Bowler}, B.~P., {Liu}, M.~C., {Shkolnik}, E.~L., \& {Dupuy}, T.~J. 2013, \apj,
  774, 55, \dodoi{10.1088/0004-637X/774/1/55}

\bibitem[{{Bowler} {et~al.}(2017{\natexlab{a}}){Bowler}, {Kraus}, {Bryan},
  {Knutson}, {Brogi}, {Rizzuto}, {Mace}, {Vanderburg}, {Liu}, {Hillenbrand}, \&
  {Cieza}}]{2017AJ....154..165B}
{Bowler}, B.~P., {Kraus}, A.~L., {Bryan}, M.~L., {et~al.} 2017{\natexlab{a}},
  \aj, 154, 165, \dodoi{10.3847/1538-3881/aa88bd}

\bibitem[{{Bowler} {et~al.}(2017{\natexlab{b}}){Bowler}, {Liu}, {Mawet}, {Ngo},
  {Malo}, {Mace}, {McLane}, {Lu}, {Tristan}, {Hinkley}, {Hillenbrand},
  {Shkolnik}, {Benneke}, \& {Best}}]{2017AJ....153...18B}
{Bowler}, B.~P., {Liu}, M.~C., {Mawet}, D., {et~al.} 2017{\natexlab{b}}, \aj,
  153, 18, \dodoi{10.3847/1538-3881/153/1/18}

\bibitem[{Buchhave {et~al.}(2018)Buchhave, Bitsch, Johansen, Latham, Bizzarro,
  Bieryla, \& Kipping}]{buc18}
Buchhave, L.~A., Bitsch, B., Johansen, A., {et~al.} 2018, The Astrophysical
  Journal, 856, 37, \dodoi{10.3847/1538-4357/aaafca}

\bibitem[{{Buchhave} {et~al.}(2014){Buchhave}, {Bizzarro}, {Latham},
  {Sasselov}, {Cochran}, {Endl}, {Isaacson}, {Juncher}, \& {Marcy}}]{buc14}
{Buchhave}, L.~A., {Bizzarro}, M., {Latham}, D.~W., {et~al.} 2014, \nat, 509,
  593, \dodoi{10.1038/nature13254}

\bibitem[{{Burgasser} {et~al.}(2010){Burgasser}, {Simcoe}, {Bochanski},
  {Saumon}, {Mamajek}, {Cushing}, {Marley}, {McMurtry}, {Pipher}, \&
  {Forrest}}]{2010ApJ...725.1405B}
{Burgasser}, A.~J., {Simcoe}, R.~A., {Bochanski}, J.~J., {et~al.} 2010, \apj,
  725, 1405, \dodoi{10.1088/0004-637X/725/2/1405}

\bibitem[{Burrows {et~al.}(1997)Burrows, Marley, Hubbard, Lunine, Guillot,
  Saumon, Freedman, Sudarsky, \& Sharp}]{bur97}
Burrows, A., Marley, M., Hubbard, W.~B., {et~al.} 1997, The Astrophysical
  Journal, 491, 856, \dodoi{10.1086/305002}

\bibitem[{{Cai} {et~al.}(2006){Cai}, {Durisen}, {Michael}, {Boley},
  {Mej{\'\i}a}, {Pickett}, \& {D'Alessio}}]{2006ApJ...636L.149C}
{Cai}, K., {Durisen}, R.~H., {Michael}, S., {et~al.} 2006, \apjl, 636, L149,
  \dodoi{10.1086/500083}

\bibitem[{{Calvet} \& {Gullbring}(1998)}]{1998ApJ...509..802C}
{Calvet}, N., \& {Gullbring}, E. 1998, \apj, 509, 802, \dodoi{10.1086/306527}

\bibitem[{{Castelli} \& {Kurucz}(2003)}]{2003IAUS..210P.A20C}
{Castelli}, F., \& {Kurucz}, R.~L. 2003, in IAU Symposium, Vol. 210, Modelling
  of Stellar Atmospheres, ed. N.~{Piskunov}, W.~W. {Weiss}, \& D.~F. {Gray},
  A20.
\newblock \doarXiv{astro-ph/0405087}

\bibitem[{{Chauvin} {et~al.}(2005){Chauvin}, {Lagrange}, {Zuckerman}, {Dumas},
  {Mouillet}, {Song}, {Beuzit}, {Lowrance}, \& {Bessell}}]{tep1}
{Chauvin}, G., {Lagrange}, A.~M., {Zuckerman}, B., {et~al.} 2005, \aap, 438,
  L29, \dodoi{10.1051/0004-6361:200500111}

\bibitem[{{Chauvin} {et~al.}(2017{\natexlab{a}}){Chauvin}, {Desidera},
  {Lagrange}, {Vigan}, {Gratton}, {Langlois}, {Bonnefoy}, {Beuzit}, {Feldt},
  {Mouillet}, {Meyer}, {Cheetham}, {Biller}, {Boccaletti}, {D'Orazi},
  {Galicher}, {Hagelberg}, {Maire}, {Mesa}, {Olofsson}, {Samland}, {Schmidt},
  {Sissa}, {Bonavita}, {Charnay}, {Cudel}, {Daemgen}, {Delorme},
  {Janin-Potiron}, {Janson}, {Keppler}, {Le Coroller}, {Ligi}, {Marleau},
  {Messina}, {Molli{\`e}re}, {Mordasini}, {M{\"u}ller}, {Peretti}, {Perrot},
  {Rodet}, {Rouan}, {Zurlo}, {Dominik}, {Henning}, {Menard}, {Schmid},
  {Turatto}, {Udry}, {Vakili}, {Abe}, {Antichi}, {Baruffolo}, {Baudoz},
  {Baudrand}, {Blanchard}, {Bazzon}, {Buey}, {Carbillet}, {Carle}, {Charton},
  {Cascone}, {Claudi}, {Costille}, {Deboulbe}, {De Caprio}, {Dohlen},
  {Fantinel}, {Feautrier}, {Fusco}, {Gigan}, {Giro}, {Gisler}, {Gluck},
  {Hubin}, {Hugot}, {Jaquet}, {Kasper}, {Madec}, {Magnard}, {Martinez},
  {Maurel}, {Le Mignant}, {M{\"o}ller-Nilsson}, {Llored}, {Moulin},
  {Orign{\'e}}, {Pavlov}, {Perret}, {Petit}, {Pragt}, {Puget}, {Rabou},
  {Ramos}, {Rigal}, {Rochat}, {Roelfsema}, {Rousset}, {Roux}, {Salasnich},
  {Sauvage}, {Sevin}, {Soenke}, {Stadler}, {Suarez}, {Weber}, {Wildi},
  {Antoniucci}, {Augereau}, {Baudino}, {Brandner}, {Engler}, {Girard}, {Gry},
  {Kral}, {Kopytova}, {Lagadec}, {Milli}, {Moutou}, {Schlieder},
  {Szul{\'a}gyi}, {Thalmann}, \& {Wahhaj}}]{CD}
{Chauvin}, G., {Desidera}, S., {Lagrange}, A.~M., {et~al.} 2017{\natexlab{a}},
  \aap, 605, L9, \dodoi{10.1051/0004-6361/201731152}

\bibitem[{{Chauvin} {et~al.}(2017{\natexlab{b}}){Chauvin}, {Desidera},
  {Lagrange}, {Vigan}, {Gratton}, {Langlois}, {Bonnefoy}, {Beuzit}, {Feldt},
  {Mouillet}, {Meyer}, {Cheetham}, {Biller}, {Boccaletti}, {D'Orazi},
  {Galicher}, {Hagelberg}, {Maire}, {Mesa}, {Olofsson}, {Samland}, {Schmidt},
  {Sissa}, {Bonavita}, {Charnay}, {Cudel}, {Daemgen}, {Delorme},
  {Janin-Potiron}, {Janson}, {Keppler}, {Le Coroller}, {Ligi}, {Marleau},
  {Messina}, {Molli{\`e}re}, {Mordasini}, {M{\"u}ller}, {Peretti}, {Perrot},
  {Rodet}, {Rouan}, {Zurlo}, {Dominik}, {Henning}, {Menard}, {Schmid},
  {Turatto}, {Udry}, {Vakili}, {Abe}, {Antichi}, {Baruffolo}, {Baudoz},
  {Baudrand}, {Blanchard}, {Bazzon}, {Buey}, {Carbillet}, {Carle}, {Charton},
  {Cascone}, {Claudi}, {Costille}, {Deboulbe}, {De Caprio}, {Dohlen},
  {Fantinel}, {Feautrier}, {Fusco}, {Gigan}, {Giro}, {Gisler}, {Gluck},
  {Hubin}, {Hugot}, {Jaquet}, {Kasper}, {Madec}, {Magnard}, {Martinez},
  {Maurel}, {Le Mignant}, {M{\"o}ller-Nilsson}, {Llored}, {Moulin},
  {Orign{\'e}}, {Pavlov}, {Perret}, {Petit}, {Pragt}, {Puget}, {Rabou},
  {Ramos}, {Rigal}, {Rochat}, {Roelfsema}, {Rousset}, {Roux}, {Salasnich},
  {Sauvage}, {Sevin}, {Soenke}, {Stadler}, {Suarez}, {Weber}, {Wildi},
  {Antoniucci}, {Augereau}, {Baudino}, {Brandner}, {Engler}, {Girard}, {Gry},
  {Kral}, {Kopytova}, {Lagadec}, {Milli}, {Moutou}, {Schlieder},
  {Szul{\'a}gyi}, {Thalmann}, \& {Wahhaj}}]{2017A...605L...9C}
---. 2017{\natexlab{b}}, \aap, 605, L9, \dodoi{10.1051/0004-6361/201731152}

\bibitem[{{Chen} {et~al.}(2011){Chen}, {Mamajek}, {Bitner}, {Pecaut}, {Su}, \&
  {Weinberger}}]{2011ApJ...738..122C}
{Chen}, C.~H., {Mamajek}, E.~E., {Bitner}, M.~A., {et~al.} 2011, \apj, 738,
  122, \dodoi{10.1088/0004-637X/738/2/122}

\bibitem[{{Chinchilla} {et~al.}(2020){Chinchilla}, {B{\'e}jar}, {Lodieu},
  {Gauza}, {Zapatero Osorio}, {Rebolo}, {Garrido}, {Alvarez}, \&
  {Manjavacas}}]{CP}
{Chinchilla}, P., {B{\'e}jar}, V. J.~S., {Lodieu}, N., {et~al.} 2020, \aap,
  633, A152, \dodoi{10.1051/0004-6361/201936130}

\bibitem[{{Choi} {et~al.}(2016){Choi}, {Dotter}, {Conroy}, {Cantiello},
  {Paxton}, \& {Johnson}}]{2016ApJ...823..102C}
{Choi}, J., {Dotter}, A., {Conroy}, C., {et~al.} 2016, \apj, 823, 102,
  \dodoi{10.3847/0004-637X/823/2/102}

\bibitem[{{Close} {et~al.}(2007){Close}, {Zuckerman}, {Song}, {Barman},
  {Marois}, {Rice}, {Siegler}, {Macintosh}, {Becklin}, {Campbell}, {Lyke},
  {Conrad}, \& {Le Mignant}}]{2007ApJ...660.1492C}
{Close}, L.~M., {Zuckerman}, B., {Song}, I., {et~al.} 2007, \apj, 660, 1492,
  \dodoi{10.1086/513417}

\bibitem[{Cumming(2004)}]{cum04}
Cumming, A. 2004, Monthly Notices of the Royal Astronomical Society, 354, 1165,
  \dodoi{10.1111/j.1365-2966.2004.08275.x}

\bibitem[{{Currie} {et~al.}(2014){Currie}, {Burrows}, \&
  {Daemgen}}]{2014ApJ...787..104C}
{Currie}, T., {Burrows}, A., \& {Daemgen}, S. 2014, \apj, 787, 104,
  \dodoi{10.1088/0004-637X/787/2/104}

\bibitem[{{D'Angelo} \& {Lissauer}(2018)}]{2018haex.bookE.140D}
{D'Angelo}, G., \& {Lissauer}, J.~J. 2018, {Formation of Giant Planets}, 140,
  \dodoi{10.1007/978-3-319-55333-7_140}

\bibitem[{{Delorme} {et~al.}(2013){Delorme}, {Gagn{\'e}}, {Girard}, {Lagrange},
  {Chauvin}, {Naud}, {Lafreni{\`e}re}, {Doyon}, {Riedel}, {Bonnefoy}, \&
  {Malo}}]{Del13}
{Delorme}, P., {Gagn{\'e}}, J., {Girard}, J.~H., {et~al.} 2013, \aap, 553, L5,
  \dodoi{10.1051/0004-6361/201321169}

\bibitem[{{Ducourant} {et~al.}(2008){Ducourant}, {Teixeira}, {Chauvin},
  {Daigne}, {Le Campion}, {Song}, \& {Zuckerman}}]{DT}
{Ducourant}, C., {Teixeira}, R., {Chauvin}, G., {et~al.} 2008, \aap, 477, L1,
  \dodoi{10.1051/0004-6361:20078886}

\bibitem[{{Erspamer} \& {North}(2002)}]{2002A&A...383..227E}
{Erspamer}, D., \& {North}, P. 2002, \aap, 383, 227,
  \dodoi{10.1051/0004-6361:20011739}

\bibitem[{Faherty {et~al.}(2009)Faherty, Burgasser, West, Bochanski, Cruz,
  Shara, \& Walter}]{Faherty_2009}
Faherty, J.~K., Burgasser, A.~J., West, A.~A., {et~al.} 2009, The Astronomical
  Journal, 139, 176, \dodoi{10.1088/0004-6256/139/1/176}

\bibitem[{{Fischer} \& {Valenti}(2005)}]{fis05}
{Fischer}, D.~A., \& {Valenti}, J. 2005, \apj, 622, 1102,
  \dodoi{10.1086/428383}

\bibitem[{{Foreman-Mackey} {et~al.}(2013){Foreman-Mackey}, {Hogg}, {Lang}, \&
  {Goodman}}]{2013PASP..125..306F}
{Foreman-Mackey}, D., {Hogg}, D.~W., {Lang}, D., \& {Goodman}, J. 2013, \pasp,
  125, 306, \dodoi{10.1086/670067}

\bibitem[{{Fulton} \& {Petigura}(2018)}]{2018AJ....156..264F}
{Fulton}, B.~J., \& {Petigura}, E.~A. 2018, \aj, 156, 264,
  \dodoi{10.3847/1538-3881/aae828}

\bibitem[{{Gaia Collaboration} {et~al.}(2018){Gaia Collaboration}, {Brown},
  {Vallenari}, {Prusti}, {de Bruijne}, {Babusiaux}, {Bailer-Jones}, {Biermann},
  {Evans}, {Eyer}, {Jansen}, {Jordi}, {Klioner}, {Lammers}, {Lindegren},
  {Luri}, {Mignard}, {Panem}, {Pourbaix}, {Randich}, {Sartoretti}, {Siddiqui},
  {Soubiran}, {van Leeuwen}, {Walton}, {Arenou}, {Bastian}, {Cropper},
  {Drimmel}, {Katz}, {Lattanzi}, {Bakker}, {Cacciari}, {Casta{\~n}eda},
  {Chaoul}, {Cheek}, {De Angeli}, {Fabricius}, {Guerra}, {Holl}, {Masana},
  {Messineo}, {Mowlavi}, {Nienartowicz}, {Panuzzo}, {Portell}, {Riello},
  {Seabroke}, {Tanga}, {Th{\'e}venin}, {Gracia-Abril}, {Comoretto},
  {Garcia-Reinaldos}, {Teyssier}, {Altmann}, {Andrae}, {Audard},
  {Bellas-Velidis}, {Benson}, {Berthier}, {Blomme}, {Burgess}, {Busso},
  {Carry}, {Cellino}, {Clementini}, {Clotet}, {Creevey}, {Davidson}, {De
  Ridder}, {Delchambre}, {Dell'Oro}, {Ducourant},
  {Fern{\'a}ndez-Hern{\'a}ndez}, {Fouesneau}, {Fr{\'e}mat}, {Galluccio},
  {Garc{\'\i}a-Torres}, {Gonz{\'a}lez-N{\'u}{\~n}ez}, {Gonz{\'a}lez-Vidal},
  {Gosset}, {Guy}, {Halbwachs}, {Hambly}, {Harrison}, {Hern{\'a}ndez},
  {Hestroffer}, {Hodgkin}, {Hutton}, {Jasniewicz}, {Jean-Antoine-Piccolo},
  {Jordan}, {Korn}, {Krone-Martins}, {Lanzafame}, {Lebzelter}, {L{\"o}ffler},
  {Manteiga}, {Marrese}, {Mart{\'\i}n-Fleitas}, {Moitinho}, {Mora}, {Muinonen},
  {Osinde}, {Pancino}, {Pauwels}, {Petit}, {Recio-Blanco}, {Richards},
  {Rimoldini}, {Robin}, {Sarro}, {Siopis}, {Smith}, {Sozzetti}, {S{\"u}veges},
  {Torra}, {van Reeven}, {Abbas}, {Abreu Aramburu}, {Accart}, {Aerts},
  {Altavilla}, {{\'A}lvarez}, {Alvarez}, {Alves}, {Anderson}, {Andrei},
  {Anglada Varela}, {Antiche}, {Antoja}, {Arcay}, {Astraatmadja}, {Bach},
  {Baker}, {Balaguer-N{\'u}{\~n}ez}, {Balm}, {Barache}, {Barata}, {Barbato},
  {Barblan}, {Barklem}, {Barrado}, {Barros}, {Barstow}, {Bartholom{\'e}
  Mu{\~n}oz}, {Bassilana}, {Becciani}, {Bellazzini}, {Berihuete}, {Bertone},
  {Bianchi}, {Bienaym{\'e}}, {Blanco-Cuaresma}, {Boch}, {Boeche}, {Bombrun},
  {Borrachero}, {Bossini}, {Bouquillon}, {Bourda}, {Bragaglia}, {Bramante},
  {Breddels}, {Bressan}, {Brouillet}, {Br{\"u}semeister}, {Brugaletta},
  {Bucciarelli}, {Burlacu}, {Busonero}, {Butkevich}, {Buzzi}, {Caffau},
  {Cancelliere}, {Cannizzaro}, {Cantat-Gaudin}, {Carballo}, {Carlucci},
  {Carrasco}, {Casamiquela}, {Castellani}, {Castro-Ginard}, {Charlot},
  {Chemin}, {Chiavassa}, {Cocozza}, {Costigan}, {Cowell}, {Crifo}, {Crosta},
  {Crowley}, {Cuypers}, {Dafonte}, {Damerdji}, {Dapergolas}, {David}, {David},
  {de Laverny}, {De Luise}, {De March}, {de Martino}, {de Souza}, {de Torres},
  {Debosscher}, {del Pozo}, {Delbo}, {Delgado}, {Delgado}, {Di Matteo},
  {Diakite}, {Diener}, {Distefano}, {Dolding}, {Drazinos}, {Dur{\'a}n},
  {Edvardsson}, {Enke}, {Eriksson}, {Esquej}, {Eynard Bontemps}, {Fabre},
  {Fabrizio}, {Faigler}, {Falc{\~a}o}, {Farr{\`a}s Casas}, {Federici},
  {Fedorets}, {Fernique}, {Figueras}, {Filippi}, {Findeisen}, {Fonti},
  {Fraile}, {Fraser}, {Fr{\'e}zouls}, {Gai}, {Galleti}, {Garabato},
  {Garc{\'\i}a-Sedano}, {Garofalo}, {Garralda}, {Gavel}, {Gavras}, {Gerssen},
  {Geyer}, {Giacobbe}, {Gilmore}, {Girona}, {Giuffrida}, {Glass}, {Gomes},
  {Granvik}, {Gueguen}, {Guerrier}, {Guiraud}, {Guti{\'e}rrez-S{\'a}nchez},
  {Haigron}, {Hatzidimitriou}, {Hauser}, {Haywood}, {Heiter}, {Helmi}, {Heu},
  {Hilger}, {Hobbs}, {Hofmann}, {Holland}, {Huckle}, {Hypki}, {Icardi},
  {Jan{\ss}en}, {Jevardat de Fombelle}, {Jonker}, {Juh{\'a}sz}, {Julbe},
  {Karampelas}, {Kewley}, {Klar}, {Kochoska}, {Kohley}, {Kolenberg},
  {Kontizas}, {Kontizas}, {Koposov}, {Kordopatis}, {Kostrzewa-Rutkowska},
  {Koubsky}, {Lambert}, {Lanza}, {Lasne}, {Lavigne}, {Le Fustec}, {Le
  Poncin-Lafitte}, {Lebreton}, {Leccia}, {Leclerc}, {Lecoeur-Taibi},
  {Lenhardt}, {Leroux}, {Liao}, {Licata}, {Lindstr{\o}m}, {Lister}, {Livanou},
  {Lobel}, {L{\'o}pez}, {Managau}, {Mann}, {Mantelet}, {Marchal}, {Marchant},
  {Marconi}, {Marinoni}, {Marschalk{\'o}}, {Marshall}, {Martino}, {Marton},
  {Mary}, {Massari}, {Matijevi{\v{c}}}, {Mazeh}, {McMillan}, {Messina},
  {Michalik}, {Millar}, {Molina}, {Molinaro}, {Moln{\'a}r}, {Montegriffo},
  {Mor}, {Morbidelli}, {Morel}, {Morris}, {Mulone}, {Muraveva}, {Musella},
  {Nelemans}, {Nicastro}, {Noval}, {O'Mullane}, {Ord{\'e}novic},
  {Ord{\'o}{\~n}ez-Blanco}, {Osborne}, {Pagani}, {Pagano}, {Pailler},
  {Palacin}, {Palaversa}, {Panahi}, {Pawlak}, {Piersimoni}, {Pineau}, {Plachy},
  {Plum}, {Poggio}, {Poujoulet}, {Pr{\v{s}}a}, {Pulone}, {Racero}, {Ragaini},
  {Rambaux}, {Ramos-Lerate}, {Regibo}, {Reyl{\'e}}, {Riclet}, {Ripepi}, {Riva},
  {Rivard}, {Rixon}, {Roegiers}, {Roelens}, {Romero-G{\'o}mez}, {Rowell},
  {Royer}, {Ruiz-Dern}, {Sadowski}, {Sagrist{\`a} Sell{\'e}s}, {Sahlmann},
  {Salgado}, {Salguero}, {Sanna}, {Santana-Ros}, {Sarasso}, {Savietto},
  {Schultheis}, {Sciacca}, {Segol}, {Segovia}, {S{\'e}gransan}, {Shih},
  {Siltala}, {Silva}, {Smart}, {Smith}, {Solano}, {Solitro}, {Sordo}, {Soria
  Nieto}, {Souchay}, {Spagna}, {Spoto}, {Stampa}, {Steele},
  {Steidelm{\"u}ller}, {Stephenson}, {Stoev}, {Suess}, {Surdej}, {Szabados},
  {Szegedi-Elek}, {Tapiador}, {Taris}, {Tauran}, {Taylor}, {Teixeira},
  {Terrett}, {Teyssand ier}, {Thuillot}, {Titarenko}, {Torra Clotet}, {Turon},
  {Ulla}, {Utrilla}, {Uzzi}, {Vaillant}, {Valentini}, {Valette}, {van Elteren},
  {Van Hemelryck}, {van Leeuwen}, {Vaschetto}, {Vecchiato}, {Veljanoski},
  {Viala}, {Vicente}, {Vogt}, {von Essen}, {Voss}, {Votruba}, {Voutsinas},
  {Walmsley}, {Weiler}, {Wertz}, {Wevers}, {Wyrzykowski}, {Yoldas},
  {{\v{Z}}erjal}, {Ziaeepour}, {Zorec}, {Zschocke}, {Zucker}, {Zurbach}, \&
  {Zwitter}}]{GAIA}
{Gaia Collaboration}, {Brown}, A.~G.~A., {Vallenari}, A., {et~al.} 2018, \aap,
  616, A1, \dodoi{10.1051/0004-6361/201833051}

\bibitem[{{Gaidos} \& {Mann}(2014)}]{2014ApJ...791...54G}
{Gaidos}, E., \& {Mann}, A.~W. 2014, \apj, 791, 54,
  \dodoi{10.1088/0004-637X/791/1/54}

\bibitem[{{Gauza} {et~al.}(2015){Gauza}, {B{\'e}jar}, {P{\'e}rez-Garrido},
  {Zapatero Osorio}, {Lodieu}, {Rebolo}, {Pall{\'e}}, \&
  {Nowak}}]{2015ApJ...804...96G}
{Gauza}, B., {B{\'e}jar}, V. J.~S., {P{\'e}rez-Garrido}, A., {et~al.} 2015,
  \apj, 804, 96, \dodoi{10.1088/0004-637X/804/2/96}

\bibitem[{{Ghezzi} {et~al.}(2010){Ghezzi}, {Cunha}, {Smith}, {de Ara{\'u}jo},
  {Schuler}, \& {de la Reza}}]{ghe10}
{Ghezzi}, L., {Cunha}, K., {Smith}, V.~V., {et~al.} 2010, \apj, 720, 1290,
  \dodoi{10.1088/0004-637X/720/2/1290}

\bibitem[{Goda \& Matsuo(2019)}]{god19}
Goda, S., \& Matsuo, T. 2019, The Astrophysical Journal, 876, 23,
  \dodoi{10.3847/1538-4357/ab0f9c}

\bibitem[{{Gonzalez}(1997)}]{1997MNRAS.285..403G}
{Gonzalez}, G. 1997, \mnras, 285, 403, \dodoi{10.1093/mnras/285.2.403}

\bibitem[{{Gravity Collaboration} {et~al.}(2019){Gravity Collaboration},
  {Lacour}, {Nowak}, {Wang}, {Pfuhl}, {Eisenhauer}, {Abuter}, {Amorim},
  {Anugu}, {Benisty}, {Berger}, {Beust}, {Blind}, {Bonnefoy}, {Bonnet},
  {Bourget}, {Brandner}, {Buron}, {Collin}, {Charnay}, {Chapron}, {Cl{\'e}net},
  {Coud{\'e} Du Foresto}, {de Zeeuw}, {Deen}, {Dembet}, {Dexter}, {Duvert},
  {Eckart}, {F{\"o}rster Schreiber}, {F{\'e}dou}, {Garcia}, {Garcia Lopez},
  {Gao}, {Gendron}, {Genzel}, {Gillessen}, {Gordo}, {Greenbaum}, {Habibi},
  {Haubois}, {Hau{\ss}mann}, {Henning}, {Hippler}, {Horrobin}, {Hubert},
  {Jimenez Rosales}, {Jocou}, {Kendrew}, {Kervella}, {Kolb}, {Lagrange},
  {Lapeyr{\`e}re}, {Le Bouquin}, {L{\'e}na}, {Lippa}, {Lenzen}, {Maire},
  {Molli{\`e}re}, {Ott}, {Paumard}, {Perraut}, {Perrin}, {Pueyo}, {Rabien},
  {Ram{\'\i}rez}, {Rau}, {Rodr{\'\i}guez-Coira}, {Rousset}, {Sanchez-Bermudez},
  {Scheithauer}, {Schuhler}, {Straub}, {Straubmeier}, {Sturm}, {Tacconi},
  {Vincent}, {van Dishoeck}, {von Fellenberg}, {Wank}, {Waisberg}, {Widmann},
  {Wieprecht}, {Wiest}, {Wiezorrek}, {Woillez}, {Yazici}, {Ziegler}, \&
  {Zins}}]{GC}
{Gravity Collaboration}, {Lacour}, S., {Nowak}, M., {et~al.} 2019, \aap, 623,
  L11, \dodoi{10.1051/0004-6361/201935253}

\bibitem[{{Gray} \& {Corbally}(1994)}]{1994AJ....107..742G}
{Gray}, R.~O., \& {Corbally}, C.~J. 1994, \aj, 107, 742, \dodoi{10.1086/116893}

\bibitem[{{Haffert} {et~al.}(2019){Haffert}, {Bohn}, {de Boer}, {Snellen},
  {Brinchmann}, {Girard}, {Keller}, \& {Bacon}}]{2019NatAs...3..749H}
{Haffert}, S.~Y., {Bohn}, A.~J., {de Boer}, J., {et~al.} 2019, Nature
  Astronomy, 3, 749, \dodoi{10.1038/s41550-019-0780-5}

\bibitem[{{Hayashi} {et~al.}(1985){Hayashi}, {Nakazawa}, \&
  {Nakagawa}}]{1985prpl.conf.1100H}
{Hayashi}, C., {Nakazawa}, K., \& {Nakagawa}, Y. 1985, in Protostars and
  Planets II, ed. D.~C. {Black} \& M.~S. {Matthews}, 1100--1153

\bibitem[{{Heap} {et~al.}(1994){Heap}, {Lanz}, {Hubeny}, \&
  {Lindler}}]{1994AAS...185.4812H}
{Heap}, S.~R., {Lanz}, T., {Hubeny}, I., \& {Lindler}, D. 1994, in American
  Astronomical Society Meeting Abstracts, Vol. 185, American Astronomical
  Society Meeting Abstracts, 48.12

\bibitem[{{Hinkley} {et~al.}(2013){Hinkley}, {Pueyo}, {Faherty}, {Oppenheimer},
  {Mamajek}, {Kraus}, {Rice}, {Ireland}, {David}, {Hillenbrand}, {Vasisht},
  {Cady}, {Brenner}, {Veicht}, {Nilsson}, {Zimmerman}, {Parry}, {Beichman},
  {Dekany}, {Roberts}, {Roberts}, {Baranec}, {Crepp}, {Burruss}, {Wallace},
  {King}, {Zhai}, {Lockhart}, {Shao}, {Soummer}, {Sivaramakrishnan}, \&
  {Wilson}}]{2013ApJ...779..153H}
{Hinkley}, S., {Pueyo}, L., {Faherty}, J.~K., {et~al.} 2013, \apj, 779, 153,
  \dodoi{10.1088/0004-637X/779/2/153}

\bibitem[{Hogg \& Foreman-Mackey(2018)}]{hog18}
Hogg, D.~W., \& Foreman-Mackey, D. 2018, The Astrophysical Journal Supplement
  Series, 236, 11, \dodoi{10.3847/1538-4365/aab76e}

\bibitem[{{Houdebine}(2010)}]{2010MNRAS.407.1657H}
{Houdebine}, E.~R. 2010, \mnras, 407, 1657,
  \dodoi{10.1111/j.1365-2966.2010.16827.x}

\bibitem[{{Howard} {et~al.}(2012){Howard}, {Marcy}, {Bryson}, {Jenkins},
  {Rowe}, {Batalha}, {Borucki}, {Koch}, {Dunham}, {Gautier}, {Van Cleve},
  {Cochran}, {Latham}, {Lissauer}, {Torres}, {Brown}, {Gilliland}, {Buchhave},
  {Caldwell}, {Christensen-Dalsgaard}, {Ciardi}, {Fressin}, {Haas}, {Howell},
  {Kjeldsen}, {Seager}, {Rogers}, {Sasselov}, {Steffen}, {Basri},
  {Charbonneau}, {Christiansen}, {Clarke}, {Dupree}, {Fabrycky}, {Fischer},
  {Ford}, {Fortney}, {Tarter}, {Girouard}, {Holman}, {Johnson}, {Klaus},
  {Machalek}, {Moorhead}, {Morehead}, {Ragozzine}, {Tenenbaum}, {Twicken},
  {Quinn}, {Isaacson}, {Shporer}, {Lucas}, {Walkowicz}, {Welsh}, {Boss},
  {Devore}, {Gould}, {Smith}, {Morris}, {Prsa}, {Morton}, {Still}, {Thompson},
  {Mullally}, {Endl}, \& {MacQueen}}]{how12}
{Howard}, A.~W., {Marcy}, G.~W., {Bryson}, S.~T., {et~al.} 2012, \apjs, 201,
  15, \dodoi{10.1088/0067-0049/201/2/15}

\bibitem[{{Ida} \& {Lin}(2004)}]{Ida04}
{Ida}, S., \& {Lin}, D.~N.~C. 2004, \apj, 616, 567, \dodoi{10.1086/424830}

\bibitem[{{Ikoma} {et~al.}(2001){Ikoma}, {Emori}, \&
  {Nakazawa}}]{2001ApJ...553..999I}
{Ikoma}, M., {Emori}, H., \& {Nakazawa}, K. 2001, \apj, 553, 999,
  \dodoi{10.1086/320954}

\bibitem[{{Janson} {et~al.}(2019){Janson}, {Asensio-Torres}, {Andr{\'e}},
  {Bonnefoy}, {Delorme}, {Reffert}, {Desidera}, {Langlois}, {Chauvin},
  {Gratton}, {Bohn}, {Eriksson}, {Marleau}, {Mamajek}, {Vigan}, \&
  {Carson}}]{JM}
{Janson}, M., {Asensio-Torres}, R., {Andr{\'e}}, D., {et~al.} 2019, \aap, 626,
  A99, \dodoi{10.1051/0004-6361/201935687}

\bibitem[{{Jofr{\'e}} {et~al.}(2019){Jofr{\'e}}, {Heiter}, \&
  {Soubiran}}]{jof19}
{Jofr{\'e}}, P., {Heiter}, U., \& {Soubiran}, C. 2019, \araa, 57, 571,
  \dodoi{10.1146/annurev-astro-091918-104509}

\bibitem[{{Jofr{\'e}} {et~al.}(2014){Jofr{\'e}}, {Heiter}, {Soubiran},
  {Blanco-Cuaresma}, {Worley}, {Pancino}, {Cantat-Gaudin}, {Magrini},
  {Bergemann}, {Gonz{\'a}lez Hern{\'a}ndez}, {Hill}, {Lardo}, {de Laverny},
  {Lind}, {Masseron}, {Montes}, {Mucciarelli}, {Nordlander}, {Recio Blanco},
  {Sobeck}, {Sordo}, {Sousa}, {Tabernero}, {Vallenari}, \& {Van Eck}}]{jof14}
{Jofr{\'e}}, P., {Heiter}, U., {Soubiran}, C., {et~al.} 2014, \aap, 564, A133,
  \dodoi{10.1051/0004-6361/201322440}

\bibitem[{{Johnson} {et~al.}(2010){Johnson}, {Aller}, {Howard}, \&
  {Crepp}}]{jon10}
{Johnson}, J.~A., {Aller}, K.~M., {Howard}, A.~W., \& {Crepp}, J.~R. 2010,
  \pasp, 122, 905, \dodoi{10.1086/655775}

\bibitem[{{Johnson} {et~al.}(2017){Johnson}, {Petigura}, {Fulton}, {Marcy},
  {Howard}, {Isaacson}, {Hebb}, {Cargile}, {Morton}, {Weiss}, {Winn}, {Rogers},
  {Sinukoff}, \& {Hirsch}}]{2017AJ....154..108J}
{Johnson}, J.~A., {Petigura}, E.~A., {Fulton}, B.~J., {et~al.} 2017, \aj, 154,
  108, \dodoi{10.3847/1538-3881/aa80e7}

\bibitem[{{Kalas} {et~al.}(2008){Kalas}, {Graham}, {Chiang}, {Fitzgerald},
  {Clampin}, {Kite}, {Stapelfeldt}, {Marois}, \& {Krist}}]{2008Sci...322.1345K}
{Kalas}, P., {Graham}, J.~R., {Chiang}, E., {et~al.} 2008, Science, 322, 1345,
  \dodoi{10.1126/science.1166609}

\bibitem[{{Kenyon} \& {Hartmann}(1995)}]{1995ApJS..101..117K}
{Kenyon}, S.~J., \& {Hartmann}, L. 1995, \apjs, 101, 117,
  \dodoi{10.1086/192235}

\bibitem[{{Keppler} {et~al.}(2018){Keppler}, {Benisty}, {M{\"u}ller},
  {Henning}, {van Boekel}, {Cantalloube}, {Ginski}, {van Holstein}, {Maire},
  {Pohl}, {Samland }, {Avenhaus}, {Baudino}, {Boccaletti}, {de Boer},
  {Bonnefoy}, {Chauvin}, {Desidera}, {Langlois}, {Lazzoni}, {Marleau},
  {Mordasini}, {Pawellek}, {Stolker}, {Vigan}, {Zurlo}, {Birnstiel},
  {Brandner}, {Feldt}, {Flock}, {Girard}, {Gratton}, {Hagelberg}, {Isella},
  {Janson}, {Juhasz}, {Kemmer}, {Kral}, {Lagrange}, {Launhardt}, {Matter},
  {M{\'e}nard}, {Milli}, {Molli{\`e}re}, {Olofsson}, {P{\'e}rez}, {Pinilla},
  {Pinte}, {Quanz}, {Schmidt}, {Udry}, {Wahhaj}, {Williams}, {Buenzli},
  {Cudel}, {Dominik}, {Galicher}, {Kasper}, {Lannier}, {Mesa}, {Mouillet},
  {Peretti}, {Perrot}, {Salter}, {Sissa}, {Wildi}, {Abe}, {Antichi},
  {Augereau}, {Baruffolo}, {Baudoz}, {Bazzon}, {Beuzit}, {Blanchard}, {Brems},
  {Buey}, {De Caprio}, {Carbillet}, {Carle}, {Cascone}, {Cheetham}, {Claudi},
  {Costille}, {Delboulb{\'e}}, {Dohlen}, {Fantinel}, {Feautrier}, {Fusco},
  {Giro}, {Gluck}, {Gry}, {Hubin}, {Hugot}, {Jaquet}, {Le Mignant}, {Llored},
  {Madec}, {Magnard}, {Martinez}, {Maurel}, {Meyer}, {M{\"o}ller-Nilsson},
  {Moulin}, {Mugnier}, {Orign{\'e}}, {Pavlov}, {Perret}, {Petit}, {Pragt},
  {Puget}, {Rabou}, {Ramos}, {Rigal}, {Rochat}, {Roelfsema}, {Rousset}, {Roux},
  {Salasnich}, {Sauvage}, {Sevin}, {Soenke}, {Stadler}, {Suarez}, {Turatto}, \&
  {Weber}}]{KB}
{Keppler}, M., {Benisty}, M., {M{\"u}ller}, A., {et~al.} 2018, \aap, 617, A44,
  \dodoi{10.1051/0004-6361/201832957}

\bibitem[{Kipping \& Sandford(2016)}]{kip16}
Kipping, D.~M., \& Sandford, E. 2016, Monthly Notices of the Royal Astronomical
  Society, 463, 1323, \dodoi{10.1093/mnras/stw1926}

\bibitem[{{Konopacky} {et~al.}(2010){Konopacky}, {Ghez}, {Barman}, {Rice},
  {Bailey}, {White}, {McLean}, \& {Duch{\^e}ne}}]{2010ApJ...711.1087K}
{Konopacky}, Q.~M., {Ghez}, A.~M., {Barman}, T.~S., {et~al.} 2010, \apj, 711,
  1087, \dodoi{10.1088/0004-637X/711/2/1087}

\bibitem[{{Konopacky} {et~al.}(2016){Konopacky}, {Rameau}, {Duch{\^e}ne},
  {Filippazzo}, {Giorla Godfrey}, {Marois}, {Nielsen}, {Pueyo}, {Rafikov},
  {Rice}, {Wang}, {Ammons}, {Bailey}, {Barman}, {Bulger}, {Bruzzone},
  {Chilcote}, {Cotten}, {Dawson}, {De Rosa}, {Doyon}, {Esposito}, {Fitzgerald},
  {Follette}, {Goodsell}, {Graham}, {Greenbaum}, {Hibon}, {Hung}, {Ingraham},
  {Kalas}, {Lafreni{\`e}re}, {Larkin}, {Macintosh}, {Maire}, {Marchis},
  {Marley}, {Matthews}, {Metchev}, {Millar-Blanchaer}, {Oppenheimer}, {Palmer},
  {Patience}, {Perrin}, {Poyneer}, {Rajan}, {Rantakyr{\"o}}, {Savransky},
  {Schneider}, {Sivaramakrishnan}, {Song}, {Soummer}, {Thomas}, {Wallace},
  {Ward-Duong}, {Wiktorowicz}, \& {Wolff}}]{2016ApJ...829L...4K}
{Konopacky}, Q.~M., {Rameau}, J., {Duch{\^e}ne}, G., {et~al.} 2016, \apjl, 829,
  L4, \dodoi{10.3847/2041-8205/829/1/L4}

\bibitem[{{Kornet} {et~al.}(2005){Kornet}, {Bodenheimer}, {R{\'o}{\.z}yczka},
  \& {Stepinski}}]{Kornet05}
{Kornet}, K., {Bodenheimer}, P., {R{\'o}{\.z}yczka}, M., \& {Stepinski}, T.~F.
  2005, \aap, 430, 1133, \dodoi{10.1051/0004-6361:20041692}

\bibitem[{{Kraus} {et~al.}(2014){Kraus}, {Ireland}, {Cieza}, {Hinkley},
  {Dupuy}, {Bowler}, \& {Liu}}]{2014ApJ...781...20K}
{Kraus}, A.~L., {Ireland}, M.~J., {Cieza}, L.~A., {et~al.} 2014, \apj, 781, 20,
  \dodoi{10.1088/0004-637X/781/1/20}

\bibitem[{{Kuzuhara} {et~al.}(2011){Kuzuhara}, {Tamura}, {Ishii}, {Kudo},
  {Nishiyama}, \& {Kandori}}]{2011AJ....141..119K}
{Kuzuhara}, M., {Tamura}, M., {Ishii}, M., {et~al.} 2011, \aj, 141, 119,
  \dodoi{10.1088/0004-6256/141/4/119}

\bibitem[{{Kuzuhara} {et~al.}(2013){Kuzuhara}, {Tamura}, {Kudo}, {Janson},
  {Kand ori}, {Brandt}, {Thalmann}, {Spiegel}, {Biller}, {Carson}, {Hori},
  {Suzuki}, {Burrows}, {Henning}, {Turner}, {McElwain}, {Moro-Mart{\'\i}n},
  {Suenaga}, {Takahashi}, {Kwon}, {Lucas}, {Abe}, {Brand ner}, {Egner},
  {Feldt}, {Fujiwara}, {Goto}, {Grady}, {Guyon}, {Hashimoto}, {Hayano},
  {Hayashi}, {Hayashi}, {Hodapp}, {Ishii}, {Iye}, {Knapp}, {Matsuo}, {Mayama},
  {Miyama}, {Morino}, {Nishikawa}, {Nishimura}, {Kotani}, {Kusakabe}, {Pyo},
  {Serabyn}, {Suto}, {Takami}, {Takato}, {Terada}, {Tomono}, {Watanabe},
  {Wisniewski}, {Yamada}, {Takami}, \& {Usuda}}]{2013ApJ...774...11K}
{Kuzuhara}, M., {Tamura}, M., {Kudo}, T., {et~al.} 2013, \apj, 774, 11,
  \dodoi{10.1088/0004-637X/774/1/11}

\bibitem[{{Lachapelle} {et~al.}(2015){Lachapelle}, {Lafreni{\`e}re},
  {Gagn{\'e}}, {Jayawardhana}, {Janson}, {Helling}, \&
  {Witte}}]{2015ApJ...802...61L}
{Lachapelle}, F.-R., {Lafreni{\`e}re}, D., {Gagn{\'e}}, J., {et~al.} 2015,
  \apj, 802, 61, \dodoi{10.1088/0004-637X/802/1/61}

\bibitem[{{Lagrange}(2014)}]{lag14}
{Lagrange}, A.~M. 2014, Philosophical Transactions of the Royal Society of
  London Series A, 372, 20130090, \dodoi{10.1098/rsta.2013.0090}

\bibitem[{{Leggett} {et~al.}(2014){Leggett}, {Liu}, {Dupuy}, {Morley},
  {Marley}, \& {Saumon}}]{2014ApJ...780...62L}
{Leggett}, S.~K., {Liu}, M.~C., {Dupuy}, T.~J., {et~al.} 2014, \apj, 780, 62,
  \dodoi{10.1088/0004-637X/780/1/62}

\bibitem[{{L{\'e}pine} {et~al.}(2003){L{\'e}pine}, {Rich}, \&
  {Shara}}]{2003AJ....125.1598L}
{L{\'e}pine}, S., {Rich}, R.~M., \& {Shara}, M.~M. 2003, \aj, 125, 1598,
  \dodoi{10.1086/345972}

\bibitem[{{Liu} {et~al.}(2011){Liu}, {Delorme}, {Dupuy}, {Bowler}, {Albert},
  {Artigau}, {Reyl{\'e}}, {Forveille}, \& {Delfosse}}]{2011ApJ...740..108L}
{Liu}, M.~C., {Delorme}, P., {Dupuy}, T.~J., {et~al.} 2011, \apj, 740, 108,
  \dodoi{10.1088/0004-637X/740/2/108}

\bibitem[{{Luck}(2017)}]{2017AJ....153...21L}
{Luck}, R.~E. 2017, \aj, 153, 21, \dodoi{10.3847/1538-3881/153/1/21}

\bibitem[{{Luhman} {et~al.}(2011){Luhman}, {Burgasser}, \&
  {Bochanski}}]{2011ApJ...730L...9L}
{Luhman}, K.~L., {Burgasser}, A.~J., \& {Bochanski}, J.~J. 2011, \apjl, 730,
  L9, \dodoi{10.1088/2041-8205/730/1/L9}

\bibitem[{{Luhman} {et~al.}(2012){Luhman}, {Burgasser}, {Labb{\'e}}, {Saumon},
  {Marley}, {Bochanski}, {Monson}, \& {Persson}}]{2012ApJ...744..135L}
{Luhman}, K.~L., {Burgasser}, A.~J., {Labb{\'e}}, I., {et~al.} 2012, \apj, 744,
  135, \dodoi{10.1088/0004-637X/744/2/135}

\bibitem[{{Luhman} {et~al.}(2009){Luhman}, {Mamajek}, {Allen}, {Muench}, \&
  {Finkbeiner}}]{2009ApJ...691.1265L}
{Luhman}, K.~L., {Mamajek}, E.~E., {Allen}, P.~R., {Muench}, A.~A., \&
  {Finkbeiner}, D.~P. 2009, \apj, 691, 1265,
  \dodoi{10.1088/0004-637X/691/2/1265}

\bibitem[{{Luhman} {et~al.}(2006){Luhman}, {Wilson}, {Brandner}, {Skrutskie},
  {Nelson}, {Smith}, {Peterson}, {Cushing}, \& {Young}}]{2006ApJ...649..894L}
{Luhman}, K.~L., {Wilson}, J.~C., {Brandner}, W., {et~al.} 2006, \apj, 649,
  894, \dodoi{10.1086/506517}

\bibitem[{{Luhman} {et~al.}(2007){Luhman}, {Patten}, {Marengo}, {Schuster},
  {Hora}, {Ellis}, {Stauffer}, {Sonnett}, {Winston}, {Gutermuth}, {Megeath},
  {Backman}, {Henry}, {Werner}, \& {Fazio}}]{2007ApJ...654..570L}
{Luhman}, K.~L., {Patten}, B.~M., {Marengo}, M., {et~al.} 2007, \apj, 654, 570,
  \dodoi{10.1086/509073}

\bibitem[{Ma \& Ge(2014)}]{ma14}
Ma, B., \& Ge, J. 2014, Monthly Notices of the Royal Astronomical Society, 439,
  2781, \dodoi{10.1093/mnras/stu134}

\bibitem[{{Macintosh} {et~al.}(2015){Macintosh}, {Graham}, {Barman}, {De Rosa},
  {Konopacky}, {Marley}, {Marois}, {Nielsen}, {Pueyo}, {Rajan}, {Rameau},
  {Saumon}, {Wang}, {Patience}, {Ammons}, {Arriaga}, {Artigau}, {Beckwith},
  {Brewster}, {Bruzzone}, {Bulger}, {Burningham}, {Burrows}, {Chen}, {Chiang},
  {Chilcote}, {Dawson}, {Dong}, {Doyon}, {Draper}, {Duch{\^e}ne}, {Esposito},
  {Fabrycky}, {Fitzgerald}, {Follette}, {Fortney}, {Gerard}, {Goodsell},
  {Greenbaum}, {Hibon}, {Hinkley}, {Cotten}, {Hung}, {Ingraham},
  {Johnson-Groh}, {Kalas}, {Lafreniere}, {Larkin}, {Lee}, {Line}, {Long},
  {Maire}, {Marchis}, {Matthews}, {Max}, {Metchev}, {Millar-Blanchaer},
  {Mittal}, {Morley}, {Morzinski}, {Murray-Clay}, {Oppenheimer}, {Palmer},
  {Patel}, {Perrin}, {Poyneer}, {Rafikov}, {Rantakyr{\"o}}, {Rice}, {Rojo},
  {Rudy}, {Ruffio}, {Ruiz}, {Sadakuni}, {Saddlemyer}, {Salama}, {Savransky},
  {Schneider}, {Sivaramakrishnan}, {Song}, {Soummer}, {Thomas}, {Vasisht},
  {Wallace}, {Ward-Duong}, {Wiktorowicz}, {Wolff}, \&
  {Zuckerman}}]{2015Sci...350...64M}
{Macintosh}, B., {Graham}, J.~R., {Barman}, T., {et~al.} 2015, Science, 350,
  64, \dodoi{10.1126/science.aac5891}

\bibitem[{{Maldonado} {et~al.}(2012){Maldonado}, {Eiroa}, {Villaver},
  {Montesinos}, \& {Mora}}]{mal12}
{Maldonado}, J., {Eiroa}, C., {Villaver}, E., {Montesinos}, B., \& {Mora}, A.
  2012, \aap, 541, A40, \dodoi{10.1051/0004-6361/201218800}

\bibitem[{{Maldonado} {et~al.}(2015){Maldonado}, {Eiroa}, {Villaver},
  {Montesinos}, \& {Mora}}]{2015Mal}
---. 2015, \aap, 579, A20, \dodoi{10.1051/0004-6361/201525764}

\bibitem[{{Maldonado} {et~al.}(2019){Maldonado}, {Villaver}, {Eiroa}, \&
  {Micela}}]{Mal19}
{Maldonado}, J., {Villaver}, E., {Eiroa}, C., \& {Micela}, G. 2019, \aap, 624,
  A94, \dodoi{10.1051/0004-6361/201833827}

\bibitem[{{Malo} {et~al.}(2014{\natexlab{a}}){Malo}, {Artigau}, {Doyon},
  {Lafreni{\`e}re}, {Albert}, \& {Gagn{\'e}}}]{2014ApJ...788...81M}
{Malo}, L., {Artigau}, {\'E}., {Doyon}, R., {et~al.} 2014{\natexlab{a}}, \apj,
  788, 81, \dodoi{10.1088/0004-637X/788/1/81}

\bibitem[{{Malo} {et~al.}(2014{\natexlab{b}}){Malo}, {Doyon}, {Feiden},
  {Albert}, {Lafreni{\`e}re}, {Artigau}, {Gagn{\'e}}, \&
  {Riedel}}]{2014ApJ...792...37M}
{Malo}, L., {Doyon}, R., {Feiden}, G.~A., {et~al.} 2014{\natexlab{b}}, \apj,
  792, 37, \dodoi{10.1088/0004-637X/792/1/37}

\bibitem[{{Mamajek}(2012)}]{2012ApJ...754L..20M}
{Mamajek}, E.~E. 2012, \apjl, 754, L20, \dodoi{10.1088/2041-8205/754/2/L20}

\bibitem[{{Manoj} {et~al.}(2011){Manoj}, {Kim}, {Furlan}, {McClure}, {Luhman},
  {Watson}, {Espaillat}, {Calvet}, {Najita}, {D'Alessio}, {Adame}, {Sargent},
  {Forrest}, {Bohac}, {Green}, \& {Arnold}}]{2011ApJS..193...11M}
{Manoj}, P., {Kim}, K.~H., {Furlan}, E., {et~al.} 2011, \apjs, 193, 11,
  \dodoi{10.1088/0067-0049/193/1/11}

\bibitem[{{Marois} {et~al.}(2008){Marois}, {Macintosh}, {Barman}, {Zuckerman},
  {Song}, {Patience}, {Lafreni{\`e}re}, \& {Doyon}}]{2008Sci...322.1348M}
{Marois}, C., {Macintosh}, B., {Barman}, T., {et~al.} 2008, Science, 322, 1348,
  \dodoi{10.1126/science.1166585}

\bibitem[{{Marois} {et~al.}(2010){Marois}, {Zuckerman}, {Konopacky},
  {Macintosh}, \& {Barman}}]{2010Natur.468.1080M}
{Marois}, C., {Zuckerman}, B., {Konopacky}, Q.~M., {Macintosh}, B., \&
  {Barman}, T. 2010, \nat, 468, 1080, \dodoi{10.1038/nature09684}

\bibitem[{{Matsuo} {et~al.}(2007){Matsuo}, {Shibai}, {Ootsubo}, \&
  {Tamura}}]{2007ApJ...662.1282M}
{Matsuo}, T., {Shibai}, H., {Ootsubo}, T., \& {Tamura}, M. 2007, \apj, 662,
  1282, \dodoi{10.1086/517964}

\bibitem[{{Mayer} {et~al.}(2002){Mayer}, {Quinn}, {Wadsley}, \&
  {Stadel}}]{2002Sci...298.1756M}
{Mayer}, L., {Quinn}, T., {Wadsley}, J., \& {Stadel}, J. 2002, Science, 298,
  1756, \dodoi{10.1126/science.1077635}

\bibitem[{{Meshkat} {et~al.}(2013){Meshkat}, {Bailey}, {Rameau}, {Bonnefoy},
  {Boccaletti}, {Mamajek}, {Kenworthy}, {Chauvin}, {Lagrange}, {Su}, \&
  {Currie}}]{2013ApJ...775L..40M}
{Meshkat}, T., {Bailey}, V., {Rameau}, J., {et~al.} 2013, \apjl, 775, L40,
  \dodoi{10.1088/2041-8205/775/2/L40}

\bibitem[{{Meshkat} {et~al.}(2017){Meshkat}, {Mawet}, {Bryan}, {Hinkley},
  {Bowler}, {Stapelfeldt}, {Batygin}, {Padgett}, {Morales}, {Serabyn},
  {Christiaens}, {Brandt}, \& {Wahhaj}}]{2017AJ....154..245M}
{Meshkat}, T., {Mawet}, D., {Bryan}, M.~L., {et~al.} 2017, \aj, 154, 245,
  \dodoi{10.3847/1538-3881/aa8e9a}

\bibitem[{{Metchev} \& {Hillenbrand}(2006)}]{2006ApJ...651.1166M}
{Metchev}, S.~A., \& {Hillenbrand}, L.~A. 2006, \apj, 651, 1166,
  \dodoi{10.1086/507836}

\bibitem[{Mizuno(1980)}]{MIZ80}
Mizuno, H. 1980, Progress of Theoretical Physics, 64, 544,
  \dodoi{10.1143/PTP.64.544}

\bibitem[{{Mohanty} {et~al.}(2007){Mohanty}, {Jayawardhana}, {Hu{\'e}lamo}, \&
  {Mamajek}}]{2007ApJ...657.1064M}
{Mohanty}, S., {Jayawardhana}, R., {Hu{\'e}lamo}, N., \& {Mamajek}, E. 2007,
  \apj, 657, 1064, \dodoi{10.1086/510877}

\bibitem[{{Mo{\'o}r} {et~al.}(2013){Mo{\'o}r}, {{\'A}brah{\'a}m},
  {K{\'o}sp{\'a}l}, {Szab{\'o}}, {Apai}, {Balog}, {Csengeri}, {Grady},
  {Henning}, {Juh{\'a}sz}, {Kiss}, {Pascucci}, {Szul{\'a}gyi}, \&
  {Vavrek}}]{2013ApJ...775L..51M}
{Mo{\'o}r}, A., {{\'A}brah{\'a}m}, P., {K{\'o}sp{\'a}l}, {\'A}., {et~al.} 2013,
  \apjl, 775, L51, \dodoi{10.1088/2041-8205/775/2/L51}

\bibitem[{{Mordasini} {et~al.}(2012){Mordasini}, {Alibert}, {Benz}, {Klahr}, \&
  {Henning}}]{2012A&A...541A..97M}
{Mordasini}, C., {Alibert}, Y., {Benz}, W., {Klahr}, H., \& {Henning}, T. 2012,
  \aap, 541, A97, \dodoi{10.1051/0004-6361/201117350}

\bibitem[{{Mulders}(2018)}]{mul18}
{Mulders}, G.~D. 2018, {Planet Populations as a Function of Stellar
  Properties}, 153, \dodoi{10.1007/978-3-319-55333-7_153}

\bibitem[{{Mulders} {et~al.}(2016){Mulders}, {Pascucci}, {Apai}, {Frasca}, \&
  {Molenda-{\.Z}akowicz}}]{2016AJ....152..187M}
{Mulders}, G.~D., {Pascucci}, I., {Apai}, D., {Frasca}, A., \&
  {Molenda-{\.Z}akowicz}, J. 2016, \aj, 152, 187,
  \dodoi{10.3847/0004-6256/152/6/187}

\bibitem[{Mulders {et~al.}(2016)Mulders, Pascucci, Apai, Frasca, \&
  Molenda-{\.{Z}}akowicz}]{Mul16}
Mulders, G.~D., Pascucci, I., Apai, D., Frasca, A., \& Molenda-{\.{Z}}akowicz,
  J. 2016, The Astronomical Journal, 152, 187,
  \dodoi{10.3847/0004-6256/152/6/187}

\bibitem[{{M\"uller, A.} {et~al.}(2018){M\"uller, A.}, {Keppler, M.}, {Henning,
  Th.}, {Samland, M.}, {Chauvin, G.}, {Beust, H.}, {Maire, A.-L.},
  {Molaverdikhani, K.}, {van Boekel, R.}, {Benisty, M.}, {Boccaletti, A.},
  {Bonnefoy, M.}, {Cantalloube, F.}, {Charnay, B.}, {Baudino, J.-L.}, {Gennaro,
  M.}, {Long, Z. C.}, {Cheetham, A.}, {Desidera, S.}, {Feldt, M.}, {Fusco, T.},
  {Girard, J.}, {Gratton, R.}, {Hagelberg, J.}, {Janson, M.}, {Lagrange,
  A.-M.}, {Langlois, M.}, {Lazzoni, C.}, {Ligi, R.}, {M\'enard, F.}, {Mesa,
  D.}, {Meyer, M.}, {Molli\`ere, P.}, {Mordasini, C.}, {Moulin, T.}, {Pavlov,
  A.}, {Pawellek, N.}, {Quanz, S. P.}, {Ramos, J.}, {Rouan, D.}, {Sissa, E.},
  {Stadler, E.}, {Vigan, A.}, {Wahhaj, Z.}, {Weber, L.}, \& {Zurlo,
  A.}}]{refId0}
{M\"uller, A.}, {Keppler, M.}, {Henning, Th.}, {et~al.} 2018, A\&A, 617, L2,
  \dodoi{10.1051/0004-6361/201833584}

\bibitem[{{Narang} {et~al.}(2018){Narang}, {Manoj}, {Furlan}, {Mordasini},
  {Henning}, {Mathew}, {Banyal}, \& {Sivarani}}]{nar18}
{Narang}, M., {Manoj}, P., {Furlan}, E., {et~al.} 2018, \aj, 156, 221,
  \dodoi{10.3847/1538-3881/aae391}

\bibitem[{{Naud} {et~al.}(2014){Naud}, {Artigau}, {Malo}, {Albert}, {Doyon},
  {Lafreni{\`e}re}, {Gagn{\'e}}, {Saumon}, {Morley}, {Allard}, {Homeier},
  {Beichman}, {Gelino}, \& {Boucher}}]{2014ApJ...787....5N}
{Naud}, M.-E., {Artigau}, {\'E}., {Malo}, L., {et~al.} 2014, \apj, 787, 5,
  \dodoi{10.1088/0004-637X/787/1/5}

\bibitem[{{Neuh{\"a}user} {et~al.}(2008){Neuh{\"a}user}, {Mugrauer},
  {Seifahrt}, {Schmidt}, \& {Vogt}}]{2008Neuh}
{Neuh{\"a}user}, R., {Mugrauer}, M., {Seifahrt}, A., {Schmidt}, T.~O.~B., \&
  {Vogt}, N. 2008, \aap, 484, 281, \dodoi{10.1051/0004-6361:20078493}

\bibitem[{{Neuh{\"a}user} \& {Schmidt}(2012)}]{neu12}
{Neuh{\"a}user}, R., \& {Schmidt}, T.~O.~B. 2012, arXiv e-prints,
  arXiv:1201.3537.
\newblock \doarXiv{1201.3537}

\bibitem[{{Nguyen} {et~al.}(2012){Nguyen}, {Brandeker}, {van Kerkwijk}, \&
  {Jayawardhana}}]{2012ApJ...745..119N}
{Nguyen}, D.~C., {Brandeker}, A., {van Kerkwijk}, M.~H., \& {Jayawardhana}, R.
  2012, \apj, 745, 119, \dodoi{10.1088/0004-637X/745/2/119}

\bibitem[{Nielsen {et~al.}(2019)Nielsen, Rosa, Macintosh, Wang, Ruffio, Chiang,
  Marley, Saumon, Savransky, Ammons, Bailey, Barman, Blain, Bulger, Burrows,
  Chilcote, Cotten, Czekala, Doyon, Duch{\^{e}}ne, Esposito, Fabrycky,
  Fitzgerald, Follette, Fortney, Gerard, Goodsell, Graham, Greenbaum, Hibon,
  Hinkley, Hirsch, Hom, Hung, Dawson, Ingraham, Kalas, Konopacky, Larkin, Lee,
  Lin, Maire, Marchis, Marois, Metchev, Millar-Blanchaer, Morzinski,
  Oppenheimer, Palmer, Patience, Perrin, Poyneer, Pueyo, Rafikov, Rajan,
  Rameau, Rantakyrö, Ren, Schneider, Sivaramakrishnan, Song, Soummer, Tallis,
  Thomas, Ward-Duong, \& Wolff}]{nie19}
Nielsen, E.~L., Rosa, R. J.~D., Macintosh, B., {et~al.} 2019, The Astronomical
  Journal, 158, 13, \dodoi{10.3847/1538-3881/ab16e9}

\bibitem[{{Nissen} \& {Gustafsson}(2018)}]{nis18}
{Nissen}, P.~E., \& {Gustafsson}, B. 2018, \aapr, 26, 6,
  \dodoi{10.1007/s00159-018-0111-3}

\bibitem[{{Patience} {et~al.}(2012){Patience}, {King}, {De Rosa}, {Vigan},
  {Witte}, {Rice}, {Helling}, \& {Hauschildt}}]{PK}
{Patience}, J., {King}, R.~R., {De Rosa}, R.~J., {et~al.} 2012, \aap, 540, A85,
  \dodoi{10.1051/0004-6361/201118058}

\bibitem[{{Pecaut} {et~al.}(2012){Pecaut}, {Mamajek}, \&
  {Bubar}}]{2012ApJ...746..154P}
{Pecaut}, M.~J., {Mamajek}, E.~E., \& {Bubar}, E.~J. 2012, \apj, 746, 154,
  \dodoi{10.1088/0004-637X/746/2/154}

\bibitem[{Pedregosa {et~al.}(2011)Pedregosa, Varoquaux, Gramfort, Michel,
  Thirion, Grisel, Blondel, Prettenhofer, Weiss, Dubourg, Vanderplas, Passos,
  Cournapeau, Brucher, Perrot, \& Duchesnay}]{scikit-learn}
Pedregosa, F., Varoquaux, G., Gramfort, A., {et~al.} 2011, Journal of Machine
  Learning Research, 12, 2825

\bibitem[{{Petigura} {et~al.}(2017){Petigura}, {Howard}, {Marcy}, {Johnson},
  {Isaacson}, {Cargile}, {Hebb}, {Fulton}, {Weiss}, {Morton}, {Winn}, {Rogers},
  {Sinukoff}, {Hirsch}, \& {Crossfield}}]{2017AJ....154..107P}
{Petigura}, E.~A., {Howard}, A.~W., {Marcy}, G.~W., {et~al.} 2017, \aj, 154,
  107, \dodoi{10.3847/1538-3881/aa80de}

\bibitem[{{Petigura} {et~al.}(2018){Petigura}, {Marcy}, {Winn}, {Weiss},
  {Fulton}, {Howard}, {Sinukoff}, {Isaacson}, {Morton}, \&
  {Johnson}}]{2018AJ....155...89P}
{Petigura}, E.~A., {Marcy}, G.~W., {Winn}, J.~N., {et~al.} 2018, \aj, 155, 89,
  \dodoi{10.3847/1538-3881/aaa54c}

\bibitem[{Petigura {et~al.}(2018)Petigura, Marcy, Winn, Weiss, Fulton, Howard,
  Sinukoff, Isaacson, Morton, \& Johnson}]{pet18}
Petigura, E.~A., Marcy, G.~W., Winn, J.~N., {et~al.} 2018, The Astronomical
  Journal, 155, 89, \dodoi{10.3847/1538-3881/aaa54c}

\bibitem[{{Piskunov} {et~al.}(1995){Piskunov}, {Kupka}, {Ryabchikova}, {Weiss},
  \& {Jeffery}}]{pis95}
{Piskunov}, N.~E., {Kupka}, F., {Ryabchikova}, T.~A., {Weiss}, W.~W., \&
  {Jeffery}, C.~S. 1995, \aaps, 112, 525

\bibitem[{{Piso} \& {Youdin}(2014)}]{2014ApJ...786...21P}
{Piso}, A.-M.~A., \& {Youdin}, A.~N. 2014, \apj, 786, 21,
  \dodoi{10.1088/0004-637X/786/1/21}

\bibitem[{{Piso} {et~al.}(2015){Piso}, {Youdin}, \&
  {Murray-Clay}}]{2015ApJ...800...82P}
{Piso}, A.-M.~A., {Youdin}, A.~N., \& {Murray-Clay}, R.~A. 2015, \apj, 800, 82,
  \dodoi{10.1088/0004-637X/800/2/82}

\bibitem[{{Pollack} {et~al.}(1996){Pollack}, {Hubickyj}, {Bodenheimer},
  {Lissauer}, {Podolak}, \& {Greenzweig}}]{Pollack96}
{Pollack}, J.~B., {Hubickyj}, O., {Bodenheimer}, P., {et~al.} 1996, \icarus,
  124, 62, \dodoi{10.1006/icar.1996.0190}

\bibitem[{{Rameau} {et~al.}(2013){Rameau}, {Chauvin}, {Lagrange}, {Meshkat},
  {Boccaletti}, {Quanz}, {Currie}, {Mawet}, {Girard}, {Bonnefoy}, \&
  {Kenworthy}}]{2013ApJ...779L..26R}
{Rameau}, J., {Chauvin}, G., {Lagrange}, A.~M., {et~al.} 2013, \apjl, 779, L26,
  \dodoi{10.1088/2041-8205/779/2/L26}

\bibitem[{{Ram{\'\i}rez} {et~al.}(2009){Ram{\'\i}rez}, {Mel{\'e}ndez}, \&
  {Asplund}}]{2009Ram}
{Ram{\'\i}rez}, I., {Mel{\'e}ndez}, J., \& {Asplund}, M. 2009, \aap, 508, L17,
  \dodoi{10.1051/0004-6361/200913038}

\bibitem[{{Rodriguez} {et~al.}(2011){Rodriguez}, {Zuckerman}, {Melis}, \&
  {Song}}]{2011ApJ...732L..29R}
{Rodriguez}, D.~R., {Zuckerman}, B., {Melis}, C., \& {Song}, I. 2011, \apjl,
  732, L29, \dodoi{10.1088/2041-8205/732/2/L29}

\bibitem[{{Royer} {et~al.}(2007){Royer}, {Zorec}, \& {G{\'o}mez}}]{RZ}
{Royer}, F., {Zorec}, J., \& {G{\'o}mez}, A.~E. 2007, \aap, 463, 671,
  \dodoi{10.1051/0004-6361:20065224}

\bibitem[{{Santos} {et~al.}(2000){Santos}, {Israelian}, \& {Mayor}}]{san00}
{Santos}, N.~C., {Israelian}, G., \& {Mayor}, M. 2000, \aap, 363, 228.
\newblock \doarXiv{astro-ph/0009182}

\bibitem[{{Santos} {et~al.}(2001){Santos}, {Israelian}, \& {Mayor}}]{san01}
---. 2001, \aap, 373, 1019, \dodoi{10.1051/0004-6361:20010648}

\bibitem[{{Santos} {et~al.}(2004){Santos}, {Israelian}, \& {Mayor}}]{san04}
---. 2004, \aap, 415, 1153, \dodoi{10.1051/0004-6361:20034469}

\bibitem[{{Santos} {et~al.}(2013){Santos}, {Sousa}, {Mortier}, {Neves},
  {Adibekyan}, {Tsantaki}, {Delgado Mena}, {Bonfils}, {Israelian}, {Mayor}, \&
  {Udry}}]{san13}
{Santos}, N.~C., {Sousa}, S.~G., {Mortier}, A., {et~al.} 2013, \aap, 556, A150,
  \dodoi{10.1051/0004-6361/201321286}

\bibitem[{{Santos} {et~al.}(2017){Santos}, {Adibekyan}, {Figueira},
  {Andreasen}, {Barros}, {Delgado-Mena}, {Demangeon}, {Faria}, {Oshagh},
  {Sousa}, {Viana}, \& {Ferreira}}]{san17}
{Santos}, N.~C., {Adibekyan}, V., {Figueira}, P., {et~al.} 2017, \aap, 603,
  A30, \dodoi{10.1051/0004-6361/201730761}

\bibitem[{Saumon \& Marley(2008)}]{sau08}
Saumon, D., \& Marley, M.~S. 2008, The Astrophysical Journal, 689, 1327,
  \dodoi{10.1086/592734}

\bibitem[{{Schlaufman}(2018)}]{2018ApJ...853...37S}
{Schlaufman}, K.~C. 2018, \apj, 853, 37, \dodoi{10.3847/1538-4357/aa961c}

\bibitem[{{Schmidt} {et~al.}(2008){Schmidt}, {Neuh{\"a}user}, {Seifahrt},
  {Vogt}, {Bedalov}, {Helling}, {Witte}, \& {Hauschildt}}]{SN}
{Schmidt}, T.~O.~B., {Neuh{\"a}user}, R., {Seifahrt}, A., {et~al.} 2008, \aap,
  491, 311, \dodoi{10.1051/0004-6361:20078840}

\bibitem[{{Schneider} {et~al.}(2011){Schneider}, {Dedieu}, {Le Sidaner},
  {Savalle}, \& {Zolotukhin}}]{sch11}
{Schneider}, J., {Dedieu}, C., {Le Sidaner}, P., {Savalle}, R., \&
  {Zolotukhin}, I. 2011, \aap, 532, A79, \dodoi{10.1051/0004-6361/201116713}

\bibitem[{{Shkedy} {et~al.}(2007){Shkedy}, {Decin}, {Molenberghs}, \&
  {Aerts}}]{shk07}
{Shkedy}, Z., {Decin}, L., {Molenberghs}, G., \& {Aerts}, C. 2007, \mnras, 377,
  120, \dodoi{10.1111/j.1365-2966.2007.11508.x}

\bibitem[{{Snellen} \& {Brown}(2018{\natexlab{a}})}]{sne18}
{Snellen}, I.~A.~G., \& {Brown}, A.~G.~A. 2018{\natexlab{a}}, Nature Astronomy,
  2, 883, \dodoi{10.1038/s41550-018-0561-6}

\bibitem[{{Snellen} \& {Brown}(2018{\natexlab{b}})}]{2018NatAs...2..883S}
---. 2018{\natexlab{b}}, Nature Astronomy, 2, 883,
  \dodoi{10.1038/s41550-018-0561-6}

\bibitem[{{Sousa} {et~al.}(2018){Sousa}, {Adibekyan}, {Delgado-Mena}, {Santos},
  {Andreasen}, {Ferreira}, {Tsantaki}, {Barros}, {Demangeon}, {Israelian},
  {Faria}, {Figueira}, {Mortier}, {Brand{\~a}o}, {Montalto}, {Rojas-Ayala}, \&
  {Santerne}}]{sau18}
{Sousa}, S.~G., {Adibekyan}, V., {Delgado-Mena}, E., {et~al.} 2018, \aap, 620,
  A58, \dodoi{10.1051/0004-6361/201833350}

\bibitem[{{Spiegel} {et~al.}(2011){Spiegel}, {Burrows}, \& {Milsom}}]{spi11}
{Spiegel}, D.~S., {Burrows}, A., \& {Milsom}, J.~A. 2011, \apj, 727, 57,
  \dodoi{10.1088/0004-637X/727/1/57}

\bibitem[{{Stassun} {et~al.}(2019){Stassun}, {Oelkers}, {Paegert}, {Torres},
  {Pepper}, {De Lee}, {Collins}, {Latham}, {Muirhead}, {Chittidi},
  {Rojas-Ayala}, {Fleming}, {Rose}, {Tenenbaum}, {Ting}, {Kane}, {Barclay},
  {Bean}, {Brassuer}, {Charbonneau}, {Ge}, {Lissauer}, {Mann}, {McLean},
  {Mullally}, {Narita}, {Plavchan}, {Ricker}, {Sasselov}, {Seager}, {Sharma},
  {Shiao}, {Sozzetti}, {Stello}, {Vanderspek}, {Wallace}, \&
  {Winn}}]{2019AJ....158..138S}
{Stassun}, K.~G., {Oelkers}, R.~J., {Paegert}, M., {et~al.} 2019, \aj, 158,
  138, \dodoi{10.3847/1538-3881/ab3467}

\bibitem[{{Stempels} {et~al.}(2007){Stempels}, {Collier Cameron}, {Hebb},
  {Smalley}, \& {Frandsen}}]{2007MNRAS.379..773S}
{Stempels}, H.~C., {Collier Cameron}, A., {Hebb}, L., {Smalley}, B., \&
  {Frandsen}, S. 2007, \mnras, 379, 773,
  \dodoi{10.1111/j.1365-2966.2007.11976.x}

\bibitem[{{Stempels} \& {Piskunov}(2002)}]{2002A&A...391..595S}
{Stempels}, H.~C., \& {Piskunov}, N. 2002, \aap, 391, 595,
  \dodoi{10.1051/0004-6361:20020814}

\bibitem[{{Stempels} \& {Piskunov}(2003)}]{2003A&A...408..693S}
---. 2003, \aap, 408, 693, \dodoi{10.1051/0004-6361:20030637}

\bibitem[{{Todorov} {et~al.}(2010){Todorov}, {Luhman}, \&
  {McLeod}}]{2010ApJ...714L..84T}
{Todorov}, K., {Luhman}, K.~L., \& {McLeod}, K.~K. 2010, \apjl, 714, L84,
  \dodoi{10.1088/2041-8205/714/1/L84}

\bibitem[{{Tody}(1986)}]{1986SPIE..627..733T}
{Tody}, D. 1986, in Society of Photo-Optical Instrumentation Engineers (SPIE)
  Conference Series, Vol. 627, Instrumentation in astronomy VI, ed. D.~L.
  {Crawford}, 733, \dodoi{10.1117/12.968154}

\bibitem[{{Tody}(1993)}]{1993ASPC...52..173T}
{Tody}, D. 1993, in Astronomical Society of the Pacific Conference Series,
  Vol.~52, Astronomical Data Analysis Software and Systems II, ed. R.~J.
  {Hanisch}, R.~J.~V. {Brissenden}, \& J.~{Barnes}, 173

\bibitem[{{Torres} {et~al.}(2006){Torres}, {Quast}, {da Silva}, {de La Reza},
  {Melo}, \& {Sterzik}}]{2006Torres}
{Torres}, C.~A.~O., {Quast}, G.~R., {da Silva}, L., {et~al.} 2006, \aap, 460,
  695, \dodoi{10.1051/0004-6361:20065602}

\bibitem[{{Traub} \& {Oppenheimer}(2010)}]{tra10}
{Traub}, W.~A., \& {Oppenheimer}, B.~R. 2010, {Direct Imaging of Exoplanets},
  ed. S.~{Seager}, 111--156

\bibitem[{{Udry} \& {Santos}(2007{\natexlab{a}})}]{udr07}
{Udry}, S., \& {Santos}, N.~C. 2007{\natexlab{a}}, \araa, 45, 397,
  \dodoi{10.1146/annurev.astro.45.051806.110529}

\bibitem[{{Udry} \& {Santos}(2007{\natexlab{b}})}]{2007ARA&A..45..397U}
---. 2007{\natexlab{b}}, \araa, 45, 397,
  \dodoi{10.1146/annurev.astro.45.051806.110529}

\bibitem[{Ujjwal {et~al.}(2020)Ujjwal, Kartha, Mathew, Manoj, \&
  Narang}]{Ujjwal_2020}
Ujjwal, K., Kartha, S.~S., Mathew, B., Manoj, P., \& Narang, M. 2020, The
  Astronomical Journal, 159, 166, \dodoi{10.3847/1538-3881/ab76d6}

\bibitem[{{Vigan} {et~al.}(2017){Vigan}, {Bonavita}, {Biller}, {Forgan},
  {Rice}, {Chauvin}, {Desidera}, {Meunier}, {Delorme}, {Schlieder}, {Bonnefoy},
  {Carson}, {Covino}, {Hagelberg}, {Henning}, {Janson}, {Lagrange}, {Quanz},
  {Zurlo}, {Beuzit}, {Boccaletti}, {Buenzli}, {Feldt}, {Girard}, {Gratton},
  {Kasper}, {Le Coroller}, {Mesa}, {Messina}, {Meyer}, {Montagnier},
  {Mordasini}, {Mouillet}, {Moutou}, {Reggiani}, {Segransan}, \&
  {Thalmann}}]{vig17}
{Vigan}, A., {Bonavita}, M., {Biller}, B., {et~al.} 2017, \aap, 603, A3,
  \dodoi{10.1051/0004-6361/201630133}

\bibitem[{{Viswanath} {et~al.}(2020){Viswanath}, {Narang}, {Manoj}, {Mathew},
  \& {Kartha}}]{2020AJ....159..194V}
{Viswanath}, G., {Narang}, M., {Manoj}, P., {Mathew}, B., \& {Kartha}, S.~S.
  2020, \aj, 159, 194, \dodoi{10.3847/1538-3881/ab7d3b}

\bibitem[{Wagner {et~al.}(2019)Wagner, Apai, \& Kratter}]{wag19}
Wagner, K., Apai, D., \& Kratter, K.~M. 2019, The Astrophysical Journal, 877,
  46, \dodoi{10.3847/1538-4357/ab1904}

\bibitem[{{Wang} \& {Fischer}(2015)}]{wan15}
{Wang}, J., \& {Fischer}, D.~A. 2015, \aj, 149, 14,
  \dodoi{10.1088/0004-6256/149/1/14}

\bibitem[{Wang {et~al.}(2018)Wang, Graham, Dawson, Fabrycky, Rosa, Pueyo,
  Konopacky, Macintosh, Marois, Chiang, Ammons, Arriaga, Bailey, Barman,
  Bulger, Chilcote, Cotten, Doyon, Duch{\^{e}}ne, Esposito, Fitzgerald,
  Follette, Gerard, Goodsell, Greenbaum, Hibon, Hung, Ingraham, Kalas, Larkin,
  Maire, Marchis, Marley, Metchev, Millar-Blanchaer, Nielsen, Oppenheimer,
  Palmer, Patience, Perrin, Poyneer, Rajan, Rameau, Rantakyrö, Ruffio,
  Savransky, Schneider, Sivaramakrishnan, Song, Soummer, Thomas, Wallace,
  Ward-Duong, Wiktorowicz, \& Wolff}]{wan18}
Wang, J.~J., Graham, J.~R., Dawson, R., {et~al.} 2018, The Astronomical
  Journal, 156, 192, \dodoi{10.3847/1538-3881/aae150}

\bibitem[{{Wilking} {et~al.}(2005){Wilking}, {Meyer}, {Robinson}, \&
  {Greene}}]{2005AJ....130.1733W}
{Wilking}, B.~A., {Meyer}, M.~R., {Robinson}, J.~G., \& {Greene}, T.~P. 2005,
  \aj, 130, 1733, \dodoi{10.1086/432758}

\bibitem[{{Winn} \& {Fabrycky}(2015)}]{win15}
{Winn}, J.~N., \& {Fabrycky}, D.~C. 2015, \araa, 53, 409,
  \dodoi{10.1146/annurev-astro-082214-122246}

\bibitem[{{Wyatt} {et~al.}(2007){Wyatt}, {Clarke}, \& {Greaves}}]{Wyatt07}
{Wyatt}, M.~C., {Clarke}, C.~J., \& {Greaves}, J.~S. 2007, \mnras, 380, 1737,
  \dodoi{10.1111/j.1365-2966.2007.12244.x}

\bibitem[{Zakamska {et~al.}(2011)Zakamska, Pan, \& Ford}]{zak11}
Zakamska, N.~L., Pan, M., \& Ford, E.~B. 2011, Monthly Notices of the Royal
  Astronomical Society, 410, 1895, \dodoi{10.1111/j.1365-2966.2010.17570.x}

\end{thebibliography}
\bibliographystyle{aasjournal}
\end{document}